\documentclass[11pt, a4paper]{article}
\usepackage{styleBuding}
\usepackage{bm}
\usepackage{color}
\usepackage[usenames,dvipsnames,svgnames,table]{xcolor}
\usepackage{amsmath}
\usepackage{amssymb}
\usepackage{graphicx}
\usepackage{slashed}
\usepackage{soul}
\usepackage{multirow}
\usepackage{subfigure}
\usepackage{indentfirst}
\usepackage{diagbox}
\usepackage{siunitx} 
\usepackage[utf8]{inputenc}

\pdfoutput=1
\usepackage{float}
\allowdisplaybreaks[4]

\begin{document}

\title{\boldmath Unified Interpretation of Muon g-2 anomaly, 95 GeV Diphoton, and $b\bar{b}$ Excesses in the General Next-to-Minimal Supersymmetric Standard Model}



\author{Junjie Cao$^{a,c}$,  Xinglong Jia$^c$, Jingwei Lian$^{b,c}$}
\affiliation{ $^a$ School of Physics, Zhengzhou University, Zhengzhou 450000, China}
\affiliation{ $^b$ Henan Institute of Science and Technology, Xinxiang 453003, China}
\affiliation{ $^c$ School of Physics, Henan Normal University, Xinxiang 453007, China }
\emailAdd{junjiec@alumni.itp.ac.cn}
\emailAdd{lianjw@hist.edu.cn}

\abstract{We investigate three intriguing anomalies within the framework of the General Next-to-Minimal Supersymmetric Standard Model. These anomalies include a significant deviation of the experimental results for the muon anomalous magnetic moment from its Standard Model prediction, with a confidence level of $5.1\sigma$; a joint observation by the CMS and ATLAS collaborations of a diphoton excess with a local significance of $3.1 \sigma$ in the invariant mass distribution around 95.4 GeV; and a reported excess in the $b\bar{b}$  production at LEP with a local significance of $2.3 \sigma$. Through analytical and numerical analyses, we provide unified interpretations across an extensive parameter space that remain consistent with current experimental restrictions from data on the Higgs boson at 125 GeV, B-physics measurements, dark matter observables, as well as existing searches for supersymmetry and extra Higgs bosons. We attribute the muon anomaly to loops involving muon-smuon-neutralino and muon-sneutrino-chargino interactions, while attributing the diphoton and $b \bar{b}$ excesses to the resonant production of a singlet-dominated scalar. These proposed solutions are poised for experimental tests at the high-luminosity LHC and future linear colliders.

}

\maketitle

\section{Introduction\label{sec:intro}}
The muon anomalous magnetic moment, $a_\mu \equiv (g-2)/2$, stands as one of the most precisely measured quantities in particle physics. It holds a pivotal position as a crucial low-energy observable known for its sensitivity to potential new physics. The persistent deviation between the experimentally measured value and the prediction derived from the Standard Model (SM) has been regarded as a compelling indication of physics beyond the SM (BSM). Initial reports of this anomaly surfaced from the E821 experiment conducted at Brookhaven National Laboratory (BNL)~\cite{Muong-2:2006rrc} and have since been further substantiated by the Run-1 data presented by E989 experiment at FermiLab in 2021~\cite{Muong-2:2021ojo}.
In a recent development, the FermiLab Muon $g-2$ Experiment released a new measurement based on combined Run-2 and Run-3 data, demonstrating consistency with the earlier findings from BNL and Run-1 data~\cite{Muong-2:2023cdq}. Upon comparing the updated world average $a_\mu^{\rm Exp} = 116 592 059(22) \times 10^{-11}$ with the 2020 white paper prediction $a_\mu^{\rm SM} = 116591810(43) \times 10^{-11}$~\cite{Aoyama:2020ynm}, a notable discrepancy was revealed at a confidence level of $5.1\sigma$:
\begin{equation}
\Delta a_\mu \equiv a_\mu^{\rm Exp}-a_\mu^{\rm SM} = (249 \pm 48) \times 10^{-11}.
\end{equation}
 While it is essential to acknowledge that uncertainties in the computation of leading-order hadronic contributions may contribute to the observed deviation~\cite{Colangelo:2023rqr,Bray-Ali:2023whf}~\footnote{ The potential underestimation of theoretical uncertainties has been a subject of debate, thereby impeding the establishment of a conclusive comparison of the muon $g-2$ measurement with theoretical predictions at present~\cite{Venanzoni:2023mbe, Kuberski:2023qgx, Muong-2:2024hpx}. Notably, the Budapest-Marseille-Wuppertal (BMW) collaboration conducted lattice calculations on the leading-order hadronic vacuum polarization contribution, achieving sub-percent precision as detailed in~\cite{Borsanyi:2020mff}. This effort mitigated the observed discrepancy to a margin of 1.5$\sigma$ ~\cite{Davier:2023cyp}. Additionally, the determination of dispersive relations, utilizing the latest $e^+e^- \to \pi^+\pi^-$ cross section measurements by the CMD-3 experiment, demonstrates good alignment with the combined FermiLab-BNL result, despite exhibiting a discrepancy with all previous results~\cite{CMD-3:2023alj,CMD-3:2023rfe}.}, there has been considerable interest in exploring the BSM origin of this anomaly~\cite{Athron:2021iuf,Lindner:2016bgg}.

As a highly promising BSM theory, the supersymmetric (SUSY) extension of the SM has garnered extensive attention for its exquisite structure and its ability to offer natural solutions to several longstanding puzzles within the SM, like the hierarchy problem, the unification of fundamental interactions, and the dark matter (DM) enigma~\cite{Fayet:1976cr,Haber:1984rc,Gunion:1984yn,Djouadi:2005gj,Martin:1997ns,Jungman:1995df}.
Numerous studies on low-energy SUSY theories have consistently pointed towards quantum corrections from SUSY partners of SM particles (sparticles), such as smuon-neutralino or sneutrino-chargino loop effects, as potential contributors to the observed anomaly~\cite{Moroi:1995yh,Hollik:1997vb,Czarnecki:2001pv,Stockinger:2006zn,Domingo:2008bb,Cao:2019evo,Endo:2021zal,Baum:2021qzx,Baum:2023inl}.
To explain $\Delta a_\mu$ within the SUSY framework, the associated sparticles must exhibit relatively low masses. However, such light sparticles have been subject to stringent constraints from the sparticles searches experiments at the Large Hadron Collider (LHC) and dark matter detection experiments.
In the Minimal Supersymmetric Standard Model (MSSM) which is the most economical realization of supersymmetry~\cite{Haber:1984rc, Gunion:1984yn, Djouadi:2005gj}, the lightest neutralino, when serving as the lightest SUSY particle (LSP) with a predominant Bino field component, can fully account for the measured DM relic density~\cite{Planck:2018vyg}.
Recent research on the phenomenology of the MSSM has revealed that, on the premise of explaining the muon g-2 anomaly at a 2$\sigma$ level, the Bino DM achieves the correct relic density mainly through co-annihilating with Wino-like electroweakinos and/or an muon-flavored slepton, and the Higgsino mass should be larger than about 500 GeV after considering the LHC experiments ~\cite{He:2023lgi}. This implies a tuning of $\mathcal{O}(1\%)$ for predicting the Z-boson mass~\cite{Baer:2012uy}. Furthermore, in the Next-to-Minimal Supersymmetric Standard Model (NMSSM) with a $\mathbb{Z}_3$ symmetry ($\mathbb{Z}_3$)~\cite{Ellwanger:2009dp, Maniatis:2009re}, another economical implementation of supersymmetry, the situation parallels that of the MSSM when the neutralino DM is Bino-dominated~\cite{Cao:2022htd}.
As for the Singlino-like DM scenario, the constraints from dark matter detection experiments necessitate a relatively small value for $\lambda$. This requirement prefers the DM to co-annihilate with the Higgsinos to achieve the correct relic density. Notably, this situation is associated with a small Bayesian evidence, indicating a certain level of fine-tuning within this framework~\cite{Cao:2016cnv, Cao:2018rix, Zhou:2021pit, Cao:2022htd}.
Motivated by the mounting tension between natural interpretations of the g-2 anomaly and the existing experimental constraints, we have explored the general form of the NMSSM, termed as ``GNMSSM".
This exploration involves the deliberate omission of the ad hoc $\mathbb{Z}_3$ symmetry, a strategic choice driven by its advantageous role in circumventing the tadpole problem and resolving the cosmological domain-wall problem associated with the $\mathbb{Z}_3$-NMSSM~\cite{Ellwanger:2009dp}.
Our findings indicate that the GNMSSM can establish a secluded dark matter sector capable of producing results consistent with the dark matter experimental observations. Additionally, it offers a robust explanation  for the muon g-2 anomaly across a broad parameter space that aligns well with various experimental results and naturally breaks the electroweak symmetry ~\cite{Cao:2021tuh, Cao:2022ovk, Cao:2022chy}.

In March 2023, the CMS collaboration released its latest analysis of searching for a light Higgs boson, confirming a previously reported excess at $m_{\gamma\gamma} = 95.4$ GeV with a local significance of 2.9$\sigma$ by utilizing advanced analysis techniques on data collected during the first, second, and third years of Run-2~\cite{CMS:2023yay}. Besides, the ATLAS collaboration recently also published an analysis of the full Run-2 LHC dataset (140~fb$^{-1}$) to search for diphoton signals in the mass range from 66~GeV to 110~GeV~\cite{Arcangeletti}. This analysis revealed an excess at an invariant mass around 95 GeV with a local significance of $1.7\sigma$. Notably, the renormalized diphoton production rates observed by both collaborations agree within uncertainties for the same mass value, suggesting the possibility that the excesses arose from the production of a single new particle. If confirmed, this could be the first sign of new physics in the Higgs-boson sector~\cite{Biekotter:2023oen}. After neglecting possible correlations, the combined signal strength corresponds to a 3.1$\sigma$ local excess~\cite{Biekotter:2023oen}:
\begin{equation}
\mu^{\rm exp}_{\gamma\gamma} \equiv  \mu^{\rm ATLAS + CMS}_{\gamma \gamma} = \frac{\sigma(p p \to \phi \to \gamma\gamma)}{\sigma_{\rm SM} (p p \to H_{\rm SM} \to \gamma\gamma)} = 0.24^{+ 0.09}_{-0.08}.  \label{diphoton-rate}
\end{equation}
Here, $\phi$ is a postulated non-Standard scalar with $m_\phi = 95.4~{\rm GeV}$ responsible for the diphoton excess, and $\sigma_{\rm SM}$ denotes the cross section for a hypothetical SM Higgs boson $H_{\rm SM}$ at the same mass. Furthermore, the presence of a light $\phi$ could potentially be supported by the measurement of bottom quark pair events at the Large Electron Positron (LEP) in 2006. These events provided hints towards the production of $\phi$ through the process $e^+e^- \rightarrow Z\phi \to Z (b\bar{b})$ with a local significance of $2.3\sigma$. They were characterized by an invariant mass of $b\bar{b}$ around $98$~GeV and a signal strength of $\mu_{b \bar b}^\text{exp} = 0.117 \pm 0.057$~\cite{LEPWorkingGroupforHiggsbosonsearches:2003ing,Azatov:2012bz,Cao:2016uwt}.
Considering the limited mass resolution for the dijets at LEP, there exists a possibility that the observed excess in $b \bar{b}$ events could be attributed to the same particle responsible for the diphoton excess\footnote{The purported 95 GeV scalar has also been suggested by the di-$\tau$ event surplus observed by the CMS collaboration~\cite{CMS:2022goy} and the $WW$ excess revealed in Ref.~\cite{Coloretti:2023wng}. Nevertheless, the reported signal strengths are characterized by substantial experimental uncertainties and implausibly high central values~\cite{Cao:2023gkc}. For the purpose of our analysis, we will disregard these anomalies. }.

The intriguing speculation has sparked numerous investigations into the feasibility of accommodating both excesses in BSM theories~\cite{Ashanujjaman:2023etj,Aguilar-Saavedra:2020wrj, Kundu:2019nqo, Fox:2017uwr, Belyaev:2023xnv, Azevedo:2023zkg, Benbrik:2022dja, Benbrik:2022azi,Haisch:2017gql,Biekotter:2019mib,Biekotter:2019kde,Biekotter:2020cjs,Biekotter:2021qbc,Biekotter:2021ovi,Heinemeyer:2021msz,Biekotter:2022jyr,Li:2023hsr,Biekotter:2023oen,Biekotter:2023jld, Aguilar-Saavedra:2023vpd, Banik:2023ecr,Dutta:2023cig,Sachdeva:2019hvk, Vega:2018ddp,Borah:2023hqw,Arcadi:2023smv,Ahriche:2023wkj,
Chen:2023bqr,Dev:2023kzu,Liu:2024cbr,Wang:2024bkg}. Particularly, many of these inquiries involve explorations within the SUSY framework~\cite{Fan:2013gjf, Cao:2016uwt,Biekotter:2021qbc,Heinemeyer:2018wzl,Heinemeyer:2018jcd,Beskidt:2017dil,Abdelalim:2020xfk,Hollik:2020plc,Cao:2019ofo,Biekotter:2019gtq,Biekotter:2017xmf, Choi:2019yrv, Wang:2018vxp,Domingo:2018uim,Li:2022etb,Ellwanger:2023zjc,Cao:2023gkc,Li:2023kbf,Ahriche:2023hho,Liu:2024cbr,Ellwanger:2024txc,Lian:2024smg}.
Our recent research has delved into the explanation of the observed excesses in the GNMSSM by attributing them to the resonant productions of a singlet-dominated CP-even Higgs boson~\cite{Cao:2023gkc}.  We found that the observed excesses can be consistently explained in the GNMSSM across a broad parameter space without conflicting with existing experimental constraints. These constraints include data from the 125 GeV Higgs, dark matter physics, as well as collider searches for SUSY particles and additional Higgs bosons. Considering the noteworthy implication of the anomalous magnetic moment and the merits of the GNMSSM Higgs sector, this research endeavors to present comprehensive explanations that jointly address both the muon $g-2$ and the observed excesses while ensuring consistency with the various experimental constraints.

The structure of this paper is outlined as follows. In Section~\ref{Section-Model}, we provide a concise overview of the fundamental framework of the GNMSSM. This section also discusses the SUSY contributions to $\Delta a_\mu$ and analyzes the signal rates corresponding to $\gamma\gamma$ and $b\bar{b}$ events. Moving on to Section~\ref{Section-results}, we conduct an extensive exploration of the model's parameter space and present the numerical outcomes through a combination of figures and tables. Furthermore, employing dedicated Monte Carlo simulations, we comprehensively analyze the constraints imposed by the searches for supersymmetry at the LHC. Finally, Section~\ref{conclusion} draws conclusions and offers insights based on the findings.

\section{Theoretical preliminaries}  \label{Section-Model}
\subsection{The basics of GNMSSM}
The GNMSSM incorporates the most general renormalizable couplings within its superpotential, as expressed by~\cite{Ellwanger:2009dp}
\begin{eqnarray}
 W_{\rm GNMSSM} = W_{\rm Yukawa} + \lambda \hat{S}\hat{H_u} \cdot \hat{H_d} + \frac{\kappa}{3}\hat{S}^3 + \mu \hat{H_u} \cdot \hat{H_d} + \frac{1}{2} \mu^\prime \hat{S}^2 + \xi\hat{S},  \label{Superpotential}
  \end{eqnarray}
where $W_{\rm Yukawa}$ contains the quark and lepton Yukawa terms in the MSSM superpotential, $\hat{H}_u=(\hat{H}_u^+,\hat{H}_u^0)^T$ and $\hat{H}_d=(\hat{H}_d^0,\hat{H}_d^-)^T$ represent the up- and down-type $SU(2)_L$ doublet Higgs superfields, respectively, and $\hat{S}$ denotes the singlet Higgs superfield. The dimensionless coefficients $\lambda$ and $\kappa$ parameterize the interactions of the Higgs fields, similar to those in the $\mathbb{Z}_3$-NMSSM. The bilinear mass parameters $\mu$ and $\mu^\prime$ and the singlet tadpole parameter $\xi$ account for the $\mathbb{Z}_3$-symmetry violating effects. They are advantageous for solving the tadpole problem~\cite{Ellwanger:1983mg, Ellwanger:2009dp} and the cosmological domain-wall problem of the $\mathbb{Z}_3$-NMSSM~\cite{Abel:1996cr, Kolda:1998rm, Panagiotakopoulos:1998yw}.  Note that one of these parameters can be eliminated by shifting the $\hat{S}$ field and redefining the other parameters~\cite{Ross:2011xv}. Without loss of generality, we set $\xi$ to be zero in this study. In such a scenario, the bilinear parameters may originate from an underlying discrete R symmetry, such as $Z^R_4$ or $Z^R_8$, after the SUSY breaking and naturally reside at the electroweak scale~\cite{Abel:1996cr,Lee:2010gv,Lee:2011dya,Ross:2011xv,Ross:2012nr}. These parameters can significantly influence the Higgs and DM physics of the $\mathbb{Z}_3$-NMSSM, which constitutes the focal point of this study.

The soft SUSY-breaking Lagrangian for the Higgs fields in the GNMSSM is given by
\begin{equation}
\begin{aligned}
 -\mathcal{L}_{soft} = &\Bigg[\lambda A_{\lambda} S H_u \cdot H_d + \frac{1}{3} \kappa A_{\kappa} S^3+ m_3^2 H_u\cdot H_d + \frac{1}{2} {m_S^{\prime}}^2 S^2 + \xi^\prime S + h.c.\Bigg]   \\
& + m^2_{H_u}|H_u|^2 + m^2_{H_d}|H_d|^2 + m^2_{S}|S|^2 ,
\end{aligned}
  \end{equation}
where $H_u$, $H_d$, and $S$ denote the scalar components of the Higgs superfields, and $m^2_{H_u}$, $m^2_{H_d}$, and $m^2_{S}$ are their soft-breaking masses. After solving the conditional equations to minimize the scalar potential and expressing these masses in terms of the vacuum expectation values (vevs) of the Higgs fields, namely, $\left\langle H_u^0 \right\rangle = v_u/\sqrt{2}$, $\left\langle H_d^0 \right\rangle = v_d/\sqrt{2}$, and $\left\langle S \right\rangle = v_s/\sqrt{2}$, with $v = \sqrt{v_u^2+v_d^2}\simeq 246~\mathrm{GeV}$, the Higgs sector is delineated by eleven independent parameters: $\tan{\beta} \equiv v_u/v_d$, $v_s$, the Yukawa couplings $\lambda$ and $\kappa$, the soft-breaking trilinear coefficients $A_\lambda$ and $A_\kappa$, the soft-breaking tadpole coefficient $\xi^\prime$ in dimension 3,  the bilinear mass parameters $\mu$ and $\mu^\prime$, and their soft-breaking parameters $m_3^2$ and $m_S^{\prime\ 2}$.

To clarify the properties of Higgs physics, it is customary to introduce the field combinations, specifically $H_{\rm SM} \equiv \sin\beta {\rm Re}(H_u^0) + \cos\beta {\rm Re} (H_d^0)$, $H_{\rm NSM} \equiv \cos\beta {\rm Re}(H_u^0) - \sin\beta {\rm Re}(H_d^0)$, and $A_{\rm NSM} \equiv \cos\beta {\rm Im}(H_u^0) - \sin\beta  {\rm Im}(H_d^0)$, where $H_{\rm SM}$ stands for the SM Higgs field, and $H_{\rm NSM}$ and $A_{\rm NSM}$ represent the additional doublet fields~\cite{Cao:2012fz}. The elements of the $CP$-even Higgs boson mass matrix $\mathcal{M}_S^2$ in the bases $\left(H_{\rm NSM}, H_{\rm SM}, {\rm Re}[S]\right)$ are given by~\cite{Ellwanger:2009dp,Miller:2003ay}
\begin{eqnarray}
  {\cal M}^2_{S, 11}&=& \frac{ \lambda v_s (\sqrt{2} A_\lambda + \kappa v_s + \sqrt{2} \mu^\prime ) + 2 m_3^2  }{\sin 2 \beta} + \frac{1}{2} (2 m_Z^2- \lambda^2v^2)\sin^22\beta,  \label{Mass-CP-even-Higgs} \\
  {\cal M}^2_{S, 12}&=&-\frac{1}{4}(2 m_Z^2-\lambda^2v^2)\sin4\beta, \quad {\cal M}^2_{S, 13} = -\frac{\lambda v}{\sqrt{2}} ( A_\lambda + \sqrt{2} \kappa v_s + \mu^\prime ) \cos 2 \beta, \nonumber \\
  {\cal M}^2_{S, 22}&=&m_Z^2\cos^22\beta+ \frac{1}{2} \lambda^2v^2\sin^22\beta,\nonumber  \\
  {\cal M}^2_{S, 23}&=& \frac{\lambda v}{\sqrt{2}} \left[(\sqrt{2} \lambda v_s + 2 \mu) - (A_\lambda + \sqrt{2} \kappa v_s + \mu^\prime ) \sin2\beta \right], \nonumber \\
  {\cal M}^2_{S, 33}&=& \frac{(A_\lambda + \mu^\prime) \sin 2 \beta}{2 \sqrt{2} v_s} \lambda v^2   + \frac{\kappa v_s}{\sqrt{2}} (A_\kappa +  2 \sqrt{2} \kappa v_s + 3 \mu^\prime ) - \frac{\mu}{\sqrt{2} v_s} \lambda v^2 - \frac{\sqrt{2}}{v_s} \xi^\prime, \nonumber
\end{eqnarray}
while those for the CP-odd Higgs fields in the bases $\left( A_{\rm NSM}, {\rm Im}(S)\right)$ take the following forms: 
\begin{eqnarray}
{\cal M}^2_{P,11}&=& \frac{ \lambda v_s (\sqrt{2} A_\lambda + \kappa v_s + \sqrt{2} \mu^\prime ) + 2 m_3^2  }{\sin 2 \beta}, \quad {\cal M}^2_{P,12} = \frac{\lambda v}{\sqrt{2}} ( A_\lambda - \sqrt{2} \kappa v_s - \mu^\prime ), \nonumber  \\
{\cal M}^2_{P,22}&=& \frac{(A_\lambda + 2 \sqrt{2} \kappa v_s + \mu^\prime ) \sin 2 \beta }{2 \sqrt{2} v_s} \lambda v^2  - \frac{\kappa v_s}{\sqrt{2}} (3 A_\kappa + \mu^\prime) \nonumber \\ & & - \frac{\mu}{\sqrt{2} v_s} \lambda v^2 - 2 m_S^{\prime\ 2} - \frac{\sqrt{2}}{v_s} \xi^\prime. \quad \quad \label{Mass-CP-odd-Higgs}
\end{eqnarray}
After diagonalizing $\mathcal{M}^2_{S}$ and $\mathcal{M}^2_{P}$ with unitary matrices $V$ and $U$, respectively, three $CP$-even and two $CP$-odd Higgs mass eigenstates, denoted as $h_i=\{h,H,h_{\rm s}\}$ and $a_j = \{A_H, A_s\}$, respectively, are obtained:
  \begin{eqnarray}
    h_i & = & V_{h_i}^{\rm NSM} H_{\rm NSM}+V_{h_i}^{\rm SM} H_{\rm SM}+V_{h_i}^{\rm S} Re[S], \quad \quad a_j =  U_{a_j}^{\rm NSM} A_{\rm NSM}+ U_{a_j}^{\rm S} Im [S],
    \label{Vij}
  \end{eqnarray}
where $h$ represents the SM-like Higgs boson discovered at the LHC, $H$ and $A_H$ signify heavy doublet-dominated Higgs bosons, and $h_s$ and $A_s$ are singlet-dominated scalars.
Additionally, the model predicts a pair of charged Higgs, $H^\pm = \cos \beta H_u^\pm + \sin \beta H_d^\pm$, with the squared mass given by
\begin{eqnarray}
    m^2_{H_{\pm}} &=&  \frac{ \lambda v_s (\sqrt{2} A_\lambda + \kappa v_s + \sqrt{2} \mu^\prime ) + 2 m_3^2  }{\sin 2 \beta} + m^2_W - \frac{1}{2}\lambda^2 v^2. \label{Charged Hisggs Mass}
  \end{eqnarray}
  
Under current experimental constraints, the Higgs sector exhibits the following characteristics. Firstly, the $H_{\rm NSM}$ and Re$[S]$ components in the observed Higgs boson $h$ are restricted to be less than about 10\%~\cite{ATLAS:2022vkf,CMS:2022dwd,ATLAS:2024fkg}, i.e., $\sqrt{\left (V_h^{\rm NSM} \right )^2 + \left ( V_h^{\rm S} \right )^2} \lesssim 0.1$ and $|V_h^{\rm SM}| \sim 1$. In the limit of $\tan\beta \gg 1$, the composition of $h$ primarily comes from the Re$[H_u^0]$ field, while $H$ is dominated by the Re$[H_d^0]$ component. Secondly, the $CP$-even doublet scalar $H$ exhibits a nearly degenerate mass with the $CP$-odd scalar $A_H$ and the charged Higgs bosons $H^\pm$. The searches for additional Higgs bosons at the LHC, combined with indirect constraints from $B$-physics, strongly support these bosons having significant masses, e.g., $m_{H} \gtrsim 0.5~{\rm TeV}$ \cite{ATLAS:2020zms,CMS:2022goy}. Thirdly, existing collider data allow the singlet-dominated scalar states to be moderately light and possess sizable doublet components~\cite{Cao:2013gba}. This feature does not depend on the value of $\tan \beta$ and it serves as the foundational premise of this study.

The neutralino sector in the GNMSSM consists of the Bino field $\tilde{B}$, the Wino field $\tilde{W}$, the Higgsino fields $\tilde{H}_d^0$ and $\tilde{H}_u^0$, and the Singlino field $\tilde{S}$. In the bases $\psi \equiv (\tilde{B},\tilde{W},\tilde{H}_d^0,\tilde{H}_u^0,\tilde{S})$, the symmetric neutralino mass matrix is given by~\cite{Ellwanger:2009dp}:
  \begin{equation}
    {\cal M} = \left(
    \begin{array}{ccccc}
    M_1 & 0 & -m_Z \sin \theta_W \cos \beta & m_Z \sin \theta_W \sin \beta & 0 \\
      & M_2 & m_Z \cos \theta_W \cos \beta & - m_Z \cos \theta_W \sin \beta &0 \\
    & & 0 & -\mu_{tot} & - \frac{1}{\sqrt{2}} \lambda v \sin \beta \\
    & & & 0 & -\frac{1}{\sqrt{2}} \lambda v \cos \beta \\
    & & & & m_N
    \end{array}
    \right), \label{eq:MN}
    \end{equation}
where $\theta_W$ is the weak mixing angle, and $M_1$ and $M_2$ are the soft-breaking masses of the Bino and Wino fields, respectively. The Higgsino mass $\mu_{tot}$ and the Singlino mass $m_N$ are defined as $\mu_{tot} \equiv \lambda v_s/\sqrt{2} + \mu$ and  $m_N \equiv  \sqrt{2} \kappa v_s + \mu^\prime$. Diagonalizing $\cal{M}$ by a rotation matrix $N$ results in five mass eigenstates:
\begin{eqnarray}
\tilde{\chi}_i^0 = N_{i1} \psi^0_1 +   N_{i2} \psi^0_2 +   N_{i3} \psi^0_3 +   N_{i4} \psi^0_4 +   N_{i5} \psi^0_5,
\end{eqnarray}
where $\tilde{\chi}_i^0\,(i=1,2,3,4,5)$ are labeled in a mass-ascending order, and the matrix element $N_{ij}$ parameterizes the component of the field $\psi^0_j$ in $\tilde{\chi}_i^0$. The lightest neutralino $\tilde{\chi}_1^0$ typically acts as a viable DM candidate, potentially being $\tilde{B}$-dominated or $\tilde{S}$-dominated to acquire the measured DM relic density~\cite{Baum:2017enm}. A distinct feature of the GNMSSM is that the Singlet-like scalars may play crucial roles in DM physics when $\tilde{\chi}_1^0$ is the $\tilde{S}$-dominated~\cite{Baum:2017enm,Cao:2021ljw}. Specifically, due to the singlet-doublet coupling and the self-interaction term in the superpotential, these scalars may mediate the DM annihilation and the DM-nucleon scattering and may also present themselves as final states of the annihilation.

Recognizing that the parameters $\mu$, $\mu^\prime$, $m_3^2$, $m_S^{\prime\ 2}$, and $\xi^\prime$ are not directly related to experimental measurements,
we employ the masses of the heavy doublet Higgs fields, the $CP$-even and -odd singlet Higgs fields, and the Higgsino and Singlino fields, denoted as $m_A \equiv \sqrt{{\cal M}^2_{P,11}}$, $m_B \equiv \sqrt{{\cal M}^2_{S,33}}$, $m_C \equiv \sqrt{{\cal M}^2_{P,22}}$, $\mu_{tot}$, and $m_N$, respectively, as inputs. With this new set of parameters,  the original parameters are derived by the following identities:
\begin{eqnarray}
\mu &= & \mu_{tot} - \frac{\lambda}{\sqrt{2}} v_s, \quad \mu^\prime = m_N - \sqrt{2} \kappa v_s, \quad  m^2_3 = \frac{m^2_A \sin{2\beta}}{2} - \lambda v_s (\frac{\kappa v_s}{2} + \frac{\mu^\prime}{\sqrt{2}} + \frac{A_\lambda}{\sqrt{2}}), \nonumber \\
\xi^\prime &=& \frac{v_s}{\sqrt{2}} \left [ \frac{(A_\lambda + \mu^\prime) \sin 2 \beta}{2 \sqrt{2} v_s} \lambda v^2   + \frac{\kappa v_s}{\sqrt{2}} (A_\kappa +  2 \sqrt{2} \kappa v_s + 3 \mu^\prime ) - \frac{\mu}{\sqrt{2} v_s} \lambda v^2 - m_B^2 \right ],  \nonumber\\
m_S^{\prime 2} &=& \frac{1}{2} \left[ m_B^2 - m^2_C + \lambda \kappa \sin 2\beta v^2 - 2 \sqrt{2} \kappa v_s ( A_\kappa + \frac{\kappa}{\sqrt{2}} v_s + \mu^\prime) \right].   \label{Simplify-1}
\end{eqnarray}
Then Eqs.~(\ref{Mass-CP-even-Higgs}), (\ref{Mass-CP-odd-Higgs}), and (\ref{Charged Hisggs Mass}) are rewritten as
\begin{eqnarray}
 {\cal M}^2_{S, 11}&=& m_A^2 + \frac{1}{2} (2 m_Z^2- \lambda^2v^2)\sin^22\beta, \quad {\cal M}^2_{S, 12}=-\frac{1}{4}(2 m_Z^2-\lambda^2v^2)\sin4\beta, \nonumber \\
  {\cal M}^2_{S, 13} &=& -\frac{\lambda v}{\sqrt{2}} ( A_\lambda + m_N ) \cos 2 \beta, \quad {\cal M}^2_{S, 22} = m_Z^2\cos^22\beta+ \frac{1}{2} \lambda^2v^2\sin^22\beta, \nonumber  \\
  {\cal M}^2_{S, 23}&=& \frac{\lambda v}{\sqrt{2}} \left[ 2 \mu_{tot} - (A_\lambda + m_N ) \sin2\beta \right], \quad {\cal M}^2_{S, 33} = m_B^2, \quad {\cal M}^2_{P,11} = m_A^2,  \nonumber \\
 \quad {\cal M}^2_{P,22} &=& m_C^2, \quad {\cal M}^2_{P,12} = \frac{\lambda v}{\sqrt{2}} ( A_\lambda - m_N ), \quad  m^2_{H^{\pm}} = m_A^2 + m^2_W -  \frac{1}{2}\lambda^2 v^2.  \label{New-mass-matrix}
\end{eqnarray}
In the case where $\lambda$ equals zero, the singlet field corresponds to the physical particle state and its mass is interpreted as that of the particle. Although this scenario is not directly relevant to our study, we have confirmed that utilizing the new set of parameters as inputs can significantly expedite the process of identifying solutions to the excesses compared to using the old parameter set.

\subsection{\label{DMRD}Muon g-2 }
The SUSY contributions to the muon g-2, denoted as $a^{\rm SUSY}_{\mu}$, mainly involve loops mediated by a smuon and a neutralino, as well as those containing a muon-flavored sneutrino and a  chargino~\cite{Moroi:1995yh,Domingo:2008bb,Hollik:1997vb,Martin:2001st}. To highlight the essential characteristics of $a^{\rm SUSY}_{\mu}$, we present its expression in the mass insertion approximation instead of working in the mass eigenstate basis\footnote{We verified that the difference between the $a_\mu^{\rm SUSY}$ values calculated by the mass insertion approximation and the full expression is less than $3\%$.}. In the approximation at the lowest order, the contributions of $a_\mu^{\rm SUSY}$ can be categorized into four types: WHL, BHL, BHR, and BLR, obtained from the Feynman diagrams involving $\tilde{W}^0-\tilde{H}^0-\tilde{\mu}_L$ (and $\tilde{W}^\pm-\tilde{H}^\pm-\tilde{\nu}_L$), $\tilde{B}-\tilde{H}^0-\tilde{\mu}_L$, $\tilde{B}-\tilde{H}^0-\tilde{\mu}_R$, and $\tilde{B}-\tilde{\mu}_L-\tilde{\mu}_R$ fields, respectively. Their contributions take the following forms~\cite{Athron:2015rva, Moroi:1995yh,Endo:2021zal}:
\begin{eqnarray}
a_{\mu, \rm WHL}^{\rm SUSY}
    &=&\frac{\alpha_2}{8 \pi} \frac{m_{\mu}^2 M_2 \mu \tan \beta}{m_{\tilde{\nu}_\mu}^4} \left \{ 2 f_C\left(\frac{M_2^2}{m_{\tilde{\nu}_{\mu}}^2}, \frac{\mu^2}{m_{\tilde{\nu}_{\mu}}^2} \right) - \frac{m_{\tilde{\nu}_\mu}^4}{M_{\tilde{\mu}_L}^4} f_N\left(\frac{M_2^2}{M_{\tilde{\mu}_L}^2}, \frac{\mu^2}{M_{\tilde{\mu}_L}^2} \right) \right \}\,, \quad \quad
    \label{eq:WHL} \\
a_{\mu, \rm BHL}^{\rm SUSY}
  &=& \frac{\alpha_Y}{8 \pi} \frac{m_\mu^2 M_1 \mu  \tan \beta}{M_{\tilde{\mu}_L}^4} f_N\left(\frac{M_1^2}{M_{\tilde{\mu}_L}^2}, \frac{\mu^2}{M_{\tilde{\mu}_L}^2} \right)\,,
    \label{eq:BHL} \\
a_{\mu, \rm BHR}^{\rm SUSY}
  &=& - \frac{\alpha_Y}{4\pi} \frac{m_{\mu}^2 M_1 \mu \tan \beta}{M_{\tilde{\mu}_R}^4} f_N\left(\frac{M_1^2}{M_{\tilde{\mu}_R}^2}, \frac{\mu^2}{M_{\tilde{\mu}_R}^2} \right)\,,
    \label{eq:BHR} \\
a_{\mu, \rm BLR}^{\rm SUSY}
  &=& \frac{\alpha_Y}{4\pi} \frac{m_{\mu}^2  M_1 \mu \tan \beta}{M_1^4}
    f_N\left(\frac{M_{\tilde{\mu}_L}^2}{M_1^2}, \frac{M_{\tilde{\mu}_R}^2}{M_1^2} \right)\,,
    \label{eq:BLR}
\end{eqnarray}
where $M_{\tilde{\mu}_L}$ and $M_{\tilde{\mu}_R}$ represent the soft-breaking masses for left-handed and right-handed smuon fields, respectively, at the slepton mass scale, and they are approximately equal to smuon masses.  The loop functions are given by
\begin{eqnarray}
    \label{eq:loop-aprox}
    f_C(x,y)
    &=&  \frac{5-3(x+y)+xy}{(x-1)^2(y-1)^2} - \frac{2\ln x}{(x-y)(x-1)^3}+\frac{2\ln y}{(x-y)(y-1)^3} \,,
      \\
    f_N(x,y)
    &=&
      \frac{-3+x+y+xy}{(x-1)^2(y-1)^2} + \frac{2x\ln x}{(x-y)(x-1)^3}-\frac{2y\ln y}{(x-y)(y-1)^3} \,,
\end{eqnarray}
satisfying $f_C(1,1) = 1/2$ and $f_N(1,1) = 1/6$.

The following points about $a_\mu^{\rm SUSY}$ should be noted:
\begin{itemize}
\item If all the dimensional SUSY parameters involved in $a_\mu^{\rm SUSY}$ share a common value $M_{\rm SUSY}$, $a_\mu^{\rm SUSY}$ is proportional to $m_\mu^2 \tan \beta/M_{\rm SUSY}^2$. This suggests that the muon g-2 anomaly favors a large $\tan \beta$ and a moderately low SUSY scale.
\item It was verified that the ``WHL" contribution is usually much larger than the other contributions when $\mu_{\rm tot}$ is less than 1~TeV and $\tilde{\mu}_L$ is not significantly heavier than $\tilde{\mu}_R$~\cite{Cao:2021tuh}.
\item In contrast to the ``WHL", ``BHL", and ``BHR" contributions that typically decrease monotonously with the magnitude of $|\mu_{\rm tot}|$, the "BLR" contribution shows a linear dependence on  $\mu_{\rm tot}$. Consequently, the scenario characterized by an exceptionally massive Higgsino (e.g., $\mu_{\rm tot} \gtrsim 30~{\rm TeV}$) is capable of achieving the central value of the muon g-2 anomaly, even when $M_1$, $M_{\tilde{\mu}_L}$, and $M_{\tilde{\mu}_R}$ are at the TeV scale, as discussed in Ref.~\cite{Gu:2021mjd}.  However, this situation necessitates significant fine-tuning to accurately predict $m_Z$, as noted in Ref.~\cite{Baer:2012uy}. Additionally, it may conflict with vacuum stability or unitarity constraints due to the substantially enhanced coupling strength of the $\tilde{\mu}^\ast_L \tilde{\mu}_R h$ interaction~\cite{Staub:2018vux} \footnote{The vacuum stability constraints for the $\mu$-term extended $Z_3$-NMSSM, which constitutes a subset of the GNMSSM parameter space, were recently analyzed in Ref.~\cite{Cao:2021tuh}. Additionally, an extensive discussion on unitarity constraints was presented in Ref.~\cite{Goodsell:2018tti}. These studies, along with the cited references, are helpful to enhance researchers' understanding of the constraints. }.  Therefore, in this study, we impose a limit of $|\mu_{\rm tot}| \leq 1~{\rm TeV}$ to ensure that the theory complies with these constraints, resulting in a typically less significant "BLR" contribution.

\item Two-loop (2L) contributions to $a_\mu$, including 2L corrections to SM one-loop diagrams and those to SUSY one-loop diagrams~\cite{Stockinger:2006zn}, are about $-5\%$ of the one-loop prediction~\cite{Cao:2022ovk}. Hence, these 2L contributions are neglected in this study.
    
\item When the smuon trilinear coefficient $A_\mu$ becomes comparable to or exceeds $\mu_{tot} \tan \beta$, it can significantly influence the  $\tilde{\mu}_L-\tilde{\mu}_R$ transition, thereby substantially modifying the contribution  $a_{\mu, \rm BHR}^{\rm SUSY}$. In deriving the aforementioned formulae for $a_\mu^{\rm SUSY}$, we have operated under the assumption that  $A_\mu \ll \mu_{tot} \tan \beta$ and consequently neglected the effects of $A_\mu$ throughout the calculations.
\end{itemize}

\subsection{\texorpdfstring{$\gamma\gamma$}{} and \texorpdfstring{$b\bar{b}$}{} signals} \label{Section-excess}
In the context of this study, the assumed origins of the excesses are the resonant productions of the singlet-dominated scalar $h_s$. The normalized diphoton signal strength, denoted as $\mu_{\gamma\gamma}$, in the narrow width approximation is expressed as follows:
\begin{eqnarray}
	\mu_{\gamma\gamma}|_{m_{h_s} = 95.4~{\rm GeV}} &=&
  \frac{\sigma_{\rm SUSY}(p p \to h_s)}
       {\sigma_{\rm SM}(p p \to h_s )} \times
       \frac{{\rm Br}_{\rm SUSY}(h_s \to \gamma \gamma)}
       {{\rm Br}_{\rm SM}(h_s \to \gamma \gamma)} \nonumber  \\
& \simeq &  \frac{\sigma_{\rm SUSY, ggF}(pp \to h_s)}{\sigma_{\rm SM, ggF}(pp \to h_s)} \times
       \frac{{\rm \Gamma}_{\rm SUSY}(h_s\to \gamma \gamma)}{{\rm \Gamma}_{\rm SM}(h_s\to \gamma \gamma)} \times \frac{\Gamma_{\rm SM}^{\rm tot}}{\Gamma_{\rm SUSY}^{\rm tot}}  \nonumber \\
& \simeq & \frac{{\rm \Gamma}_{\rm SUSY}(h_s\to g g)}{\Gamma_{\rm SM} (h_s\to g g)} \times
  \frac{{\rm \Gamma}_{\rm SUSY}(h_s\to \gamma \gamma)}{{\rm \Gamma}_{\rm SM}(h_s\to \gamma \gamma)}
   \times \frac{1}{{\rm R}_{\rm Width}},   \nonumber \\
& \simeq & |C_{h_s g g}|^2 \times |C_{h_s \gamma \gamma}|^2 \times  \frac{1}{{\rm R}_{\rm Width}},
  \label{muCMS}
\end{eqnarray}
where the mass of $h_s$ is fixed at $95.4~{\rm GeV}$. The production rate $\sigma( p p \to h_s)$, the decay branching ratio ${\rm Br} (h_s \to \gamma \gamma)$, and width $\Gamma$, labeled with the subscript `SUSY', refer to the predictions of the GNMSSM.
The subscript `SM' denotes quantities assuming $h_s$ to have SM couplings.
Since the gluon fusion (ggF) process is the primary contribution to the Higgs production~\cite{CMS:2023yay, Arcangeletti}, we take $\sigma_{\rm SUSY}(pp \to h_s)/\sigma_{\rm SM}(pp \to h_s) \simeq \sigma_{\rm SUSY, ggF}(pp \to h_s)/\sigma_{\rm SM, ggF}(pp \to h_s) \simeq {\rm \Gamma}_{\rm SUSY}(h_s\to g g)/\Gamma_{\rm SM} (h_s\to g g) \simeq |C_{h_s g g}|^2 $, where $C_{h_s g g}$ is the ratio of the $h_s$-gluon-gluon coupling strength to its SM prediction, denoted as $C_{h_s g g} \equiv |{\cal{A}}_{\rm SUSY}^{h_s g g}/{\cal{A}}_{\rm SM}^{h_s g g}|$. The normalized coupling strength of $h_s$ to photons, $C_{h_s \gamma \gamma}$, is similarly defined. Besides, the width $\Gamma_{\rm SUSY}^{\rm tot}$  and the ratio ${\rm R}_{\rm Width} \equiv \Gamma_{\rm SUSY}^{\rm tot}/ \Gamma_{\rm SM}^{\rm tot} $ are acquired by
\begin{eqnarray}
\Gamma_{\rm SUSY}^{\rm tot} &=& \Gamma_{\rm SUSY}(h_s \to b\bar{b}) +  \Gamma_{\rm SUSY}(h_s\to \tau \bar{\tau}) +  \Gamma_{\rm SUSY}(h_s\to c\bar{c}) + \Gamma_{\rm SUSY}(h_s\to g g ) + \cdots \nonumber \\
&=& |C_{h_s b \bar{b}}|^2 \times \Gamma_{\rm SM}(h_s \to b\bar{b}) + |C_{h_s \tau \bar{\tau}}|^2 \times \Gamma_{\rm SM}(h_s \to \tau \bar{\tau}) + |C_{h_s c \bar{c}}|^2 \times \Gamma_{\rm SM}(h_s \to c \bar{c}) \nonumber \\
& & +  |C_{h_s g \bar{g}}|^2 \times \Gamma_{\rm SM}(h_s \to g \bar{g}) + \cdots,  \nonumber \\
{\rm R}_{\rm Width} & = & |C_{h_s b \bar{b}}|^2 \times {\rm Br}_{\rm SM}(h_s \to b\bar{b}) + |C_{h_s \tau \bar{\tau}}|^2 \times {\rm Br}_{\rm SM}(h_s \to \tau \bar{\tau}) + |C_{h_s c \bar{c}}|^2 \times {\rm Br}_{\rm SM}(h_s \to c \bar{c}) \nonumber \\
& & +  |C_{h_s g \bar{g}}|^2 \times {\rm Br}_{\rm SM}(h_s \to g \bar{g}) + \cdots,  \nonumber \\
& \simeq & 0.801 \times |C_{h_s b \bar{b}}|^2 + 0.083 \times |C_{h_s \tau \bar{\tau}}|^2 + 0.041 \times |C_{h_s c \bar{c}}|^2 + 0.067 \times |C_{h_s g \bar{g}}|^2, \label{mugamma}
\end{eqnarray}
respectively, where $C_{h_s f \bar{f}}$ ($f=b$, $\tau$, $c$) are the normalized couplings of $h_s$ to the fermion pair $f \bar{f}$ defined by $C_{h_s f \bar{f}} \equiv {\cal{A}}_{\rm SUSY}^{h_s f \bar{f}}/{\cal{A}}_{\rm SM}^{h_s f \bar{f}}$, and the branching ratios in the SM were obtained by the LHC Higgs Cross Section Working Group, which included all known higher-order QCD corrections~\cite{LHCHiggsCrossSectionWorkingGroup:2013rie}.

Similarly, the normalized signal strength of the $b\bar{b}$ excess follows from
\begin{eqnarray}
	\mu_{b\bar{b}}|_{m_{h_s} = 95.4~{\rm GeV}} &=&
  \frac{\sigma_{\rm SUSY}(e^+e^-\to Z h_s)}
       {\sigma_{\rm SM}(e^+e^-\to Z h_s)} \times
       \frac{{\rm Br}_{\rm SUSY}(h_s\to b\bar{b})}
       {{\rm Br}_{\rm SM}(h_s \to b\bar{b})} \nonumber \\
  & = &
  \left|C_{h_s V V}\right|^2 \times |C_{h_s b \bar{b}}|^2 \times \frac{1}{{\rm R}_{\rm Width}}.    \label{muLEP}
\end{eqnarray}
It is noteworthy that the produced Higgs boson must be CP-even to explain the $b\bar{b}$ excess, since a CP-odd one does not couple to $Z$ boson and thus gives no contributions. By contrast, the scalar may be either CP-even or CP-odd to account for the diphoton excess.

In the GNMSSM, the coupling ${\cal{A}}_{\rm SUSY}^{h_s g g}$ is contributed by the loops mediated by quarks and squarks~\cite{King:2012tr}. The code \textsf{SPheno-4.0.5}~\cite{Porod2003SPheno,Porod2011SPheno3} calculates it by the following formula:
\begin{eqnarray}
{\cal{A}}_{\rm SUSY}^{h_s g g} &=& \sum_q {\cal{A}}_{\rm SUSY}^{h_s g g, q} + \sum_{\tilde{q}} {\cal{A}}_{\rm SUSY}^{h_s g g, \tilde{q}} \nonumber \\
 {\cal{A}}_{\rm SUSY}^{h_s g g, q} &=& \sum_q C_{h_s q \bar{q}}^{tree} \times \frac{m_q(Q)}{m_q(pole)} \times {\cal{A}}_{\rm SM}^{h_s g g, q}
\end{eqnarray}
where ${\cal{A}}_{\rm SUSY}^{h_s g g, q}$ and ${\cal{A}}_{\rm SUSY}^{h_s g g, \tilde{q}}$ represent the quark and squark contributions, respectively, to the $h_s g g$ coupling in the GNMSSM~\cite{Djouadi:2005gj}, and ${\cal{A}}_{\rm SM}^{h_s g g, q}$ denotes the quark loop contribution in the SM with its expression given in Ref.~\cite{Djouadi:2005gi}. $C_{h_s q \bar{q}}^{tree}$ is the tree-level prediction of $C_{h_s q \bar{q}}$, and $m_q (Q)$ and $m_q (pole)$ are the quark running mass at the scale $Q \simeq m_{h_s}$ and the pole mass, respectively. Besides, the code has incorporated the higher-order QCD corrections to these quantities by the formulae in Ref.~\cite{Spira:1995rr, Staub:2016dxq}. ${\cal{A}}_{\rm SUSY}^{h_s \gamma \gamma}$ is similarly obtained, except that it receives additional contributions from the W boson, charged Higgs boson, and  chargino-mediated loops~\cite{King:2012tr}. We add that the supersymmetric contributions to $C_{h_s g g}$ and $C_{h_s \gamma \gamma}$ are not crucial in this study. Specifically, given the massiveness of the squarks, their contributions to the couplings are typically a few thousandths of the contribution from the SM particles.  The  charginos' contribution to $C_{h_s \gamma \gamma}$ only reaches $1\%$ in an optimum case since $\lambda$ is minor~\cite{Choi:2012he}, and the charged Higgs's contribution is at the level of $0.001\%$.

In this study, we obtained the normalized couplings of $h_s$ to fermions, WW, and ZZ by their tree-level expressions\footnote{The potentially large SUSY-QCD and SUSY-electroweak corrections to the bottom quark Yukawa coupling are minor in this study since gluino and squarks are very massive~\cite{Carena:1999py}.}. After neglecting the difference of the running mass and the pole mass and the supersymmetric contributions, we had the following relations~\cite{Ellwanger:2009dp}:
\begin{eqnarray}
C_{h_s t \bar{t}} &=&  V_{h_s}^{\rm SM} + V_{h_s}^{\rm NSM} \cot \beta  \simeq V_{h_s}^{\rm SM}, \quad C_{h_s b \bar{b}} =  V_{h_s}^{\rm SM} - V_{h_s}^{\rm NSM} \tan \beta,  \quad C_{h_s V V} = V_{h_s}^{\rm SM},  \nonumber \\
C_{h_s c \bar{c}} &=& C_{h_s t \bar{t}}, \quad \quad C_{h_s \tau \bar{\tau}} = C_{h_s b \bar{b}}, \quad \quad C_{h_s g g} \simeq C_{h_s t \bar{t}}, \quad \quad C_{h_s \gamma \gamma} \simeq V_{h_s}^{\rm SM}, \label{hs-couplings}
\end{eqnarray}
where the rotation matrix elements $V^i_j$ were defined in Eq.~(\ref{Vij}). We concluded $V_{h_s}^{\rm SM} \simeq 0.36$ and $(V_{h_s}^{\rm SM} - V_{h_s}^{\rm NSM} \tan \beta) \simeq 0.70 \times V_{h_s}^{\rm SM} \simeq 0.25 $ (or equivalently, $V_{h_s}^{\rm NSM} \tan \beta \simeq 0.11$)  to acquire the central values of $\mu_{\gamma \gamma}$ and $\mu_{b \bar{b}}$ and the preferred branching ratios to be ${\rm Br}_{\rm SUSY} (h_s \to \gamma \gamma) \simeq 1.86 \times {\rm Br}_{\rm SM} (h_s \to \gamma \gamma) \simeq 2.58 \times 10^{-3}$ and ${\rm Br}_{\rm SUSY} (h_s \to b \bar{b}) \simeq 0.90 \times {\rm Br}_{\rm SM} (h_s \to  b \bar{b}) \simeq 72.6\%$. Alternatively, if we acquired $C_{h_s g g}$ and $C_{h_s \gamma \gamma}$ by the exact formulae, as we always did in this study, these couplings might deviate from $C_{h_s t \bar{t}}$ by $4\%$ and $11\%$, respectively. In this case, we found the central values of $\mu_{\gamma \gamma}$ and $\mu_{b \bar{b}}$ corresponded to $V_{h_s}^{\rm SM} \simeq 0.35$, $(V_{h_s}^{\rm SM} - V_{h_s}^{\rm NSM} \tan \beta) \simeq 0.81 \times V_{h_s}^{\rm SM} \simeq 0.28$, or equivalently, $V_{h_s}^{\rm NSM} \tan \beta \simeq 0.07$, ${\rm Br}_{\rm SUSY} (h_s \to \gamma \gamma) \simeq 1.77 \times {\rm Br}_{\rm SM} (h_s \to \gamma \gamma) \simeq 2.5 \times 10^{-3}$, and ${\rm Br}_{\rm SUSY} (h_s \to b \bar{b}) \simeq 0.95 \times {\rm Br}_{\rm SM} (h_s \to  b \bar{b}) \simeq 76.1\%$. These results reveal that explaining the excesses requires an appropriate $C_{h_s t \bar{t}}$ and simultaneously a relatively suppressed $C_{h_s b \bar{b}}$, which are mainly decided by the Higgs mixings $V_i^j$. Particularly, the small deviations of $C_{h_s g g}$ and $C_{h_s \gamma \gamma}$ from $C_{h_s t \bar{t}}$ can significantly reduce $V_{h_s}^{\rm NSM} \tan \beta$ needed to predict the central values of the excesses, but hardly change $V_{h_s}^{\rm SM}$. The approximations also indicate that any reduction of $V_{h_s}^{\rm NSM} \tan \beta$ can enhance $C_{h_s b \bar{b}}$, leading to the increase of $\mu_{b \bar{b}}$ and the decrease of $\mu_{\gamma \gamma}$
if $C_{h_s g g}$, $C_{h_s \gamma \gamma}$, and $C_{h_s V V}$ are fixed (note that $C_{h_s g g}$ and $C_{h_s \gamma \gamma}$ are insensitive to $C_{h_s b \bar{b}}$, and $C_{h_s V V}$ is independent of $C_{h_s b \bar{b}}$).

Furthermore, we point out that explaining the diphoton and $b \bar{b}$ excesses nontrivially relies on the parameters in the Higgs sector. Specifically, after neglecting the renormalization group running of the input parameters and the radiative corrections to ${\cal{M}}_S^2$ in Eq.~(\ref{New-mass-matrix}), one can express the eigenstate equation of $h_s$ as follows:
\begin{eqnarray}
\sum_{j={\rm NSM},{\rm SM},{\rm S}} \left [ {\cal{M}}_S^2 \right ]^i_j V_{h_s}^j = m_{h_s}^2 V_{h_s}^i.
\end{eqnarray}
Noting $m_{h_s}, m_h \ll m_A$, implying that the mixings of $H_{\rm NSM}$ with $H_{\rm SM}$ and $Re[S]$ are small, we acquire the following approximations:
\begin{eqnarray}
V_{h_s}^{\rm NSM} &\simeq &  \frac{V_{h_s}^S}{\sqrt{2}} \times \frac{\lambda v \bar{A}_\lambda \cos 2 \beta}{m_A^2}, \quad \lambda \left ( \mu_{tot} - \bar{A}_\lambda \sin \beta \cos \beta \right ) \simeq \frac{V_{h_s}^{\rm SM} V_{h_s}^{\rm S}}{\sqrt{2}} \times \frac{m_{h_s}^2 - m_h^2}{v},  \nonumber \\
m_B^2 & \simeq &  m_{h_s}^2 |V_{h_s}^S|^2 + m_h^2 |V_{h_s}^{\rm SM}|^2,  \quad {\cal{M}}_{S,22}^2 \simeq m_h^2 |V_{h_s}^{\rm S}|^2 + m_{h_s}^2 |V_{h_s}^{\rm SM}|^2, \label{Approximation-relations}
\end{eqnarray}
where $\bar{A}_\lambda$ is defined as $\bar{A}_\lambda \equiv A_\lambda + m_N$. These expressions indicate that the observed excesses have restricted $m_B$ and ${\cal{M}}_{S,22}$ within narrow ranges. They also suggest
\begin{eqnarray}
\lambda \simeq 0.06 \times \left ( \frac{V_{h_s}^{\rm SM}}{0.35} \right ) \times \left ( \frac{\mu_{tot} - \bar{A}_\lambda \sin \beta \cos \beta}{100~{\rm Gev}} \right )^{-1}, \label{Approximation-relations-1}
\end{eqnarray}
and
\begin{eqnarray}
\lambda \gtrsim 0.017 \times \frac{1}{|\cos 2\beta|} \times \left ( \frac{\tan \beta }{50} \right )^{-1} \times  \left ( \frac{\bar{A}_\lambda }{2~{\rm TeV}} \right )^{-1} \times \left ( \frac{m_A}{2~{\rm TeV}} \right )^2,  \label{Approximation-relations-2}
\end{eqnarray}
to predict $V_{h_s}^{\rm NSM} \tan \beta \gtrsim 0.07$. Given that the LHC searches for electroweakinos have established $\mu_{tot} \gtrsim 200~{\rm GeV}$~\cite{ATLAS:2021moa}, one can deduce that predicting the central values of the excesses favors $\lambda \lesssim 0.03$ under the condition $\bar{A}_\lambda \sin \beta \cos \beta \ll \mu_{tot}$. This in turn necessitates $\tan \beta/50 \times \bar{A}_\lambda \gtrsim 1.8~{\rm TeV}$ when $m_A = 2~{\rm TeV}$. These requirements can be met when $\bar{A}_\lambda$ reaches several TeV and $\tan \beta$ takes sufficiently large values. On the other hand, for moderate values of $\tan \beta$ where $\mu_{tot} - \bar{A}_\lambda \sin \beta \cos \beta \simeq 10~{\rm GeV}$ is acquired, $\lambda$ can be as large as approximately $0.5$ to account for the excesses, as demonstrated in Ref.~\cite{Ellwanger:2024txc}. However, this scenario relies on a fine-tuned cancellation between two independent terms, $\mu_{tot}$ and  $\bar{A}_\lambda \sin \beta \cos \beta$, and thus, its posterior probability density function (PDF) in Bayesian inference is suppressed.

\section{Explanations of the anomalies}  \label{Section-results}
This section introduces the sampling strategy used in this work and provides detailed numerical results by which we elucidate the characteristics of the GNMSSM in explaining the muon g-2 anomaly and the diphoton and $b\bar{b}$ excesses.  The numerical calculation process begins with utilizing the package \textsf{SARAH-4.15.3}~\cite{SARAH_Staub2008,SARAH3_Staub2012,SARAH4_Staub2013,SARAH_Staub2015} to build the model routines of the GNMSSM. Subsequently, the codes \textsf{SPheno-4.0.5}~\cite{Porod2003SPheno,Porod2011SPheno3} and \textsf{FlavorKit}~\cite{Porod:2014xia} are employed to generate particle spectrum and compute low energy flavor observables, respectively. We calculate the DM physics observables with the package \textsf{MicrOMEGAs-5.0.4}~\cite{Belanger2002,Belanger2004,Belanger2005,Belanger2006,BelangerRD2006qa,Belanger2008,Belanger2010pz,Belanger2013,Barducci2016pcb,Belanger2018}. Finally, we analyze the acquired samples by means of the posterior PDF in Bayesian inference and the profile likelihood (PL) in Frequentist statistics~\cite{Fowlie:2016hew}. We also discuss the impacts of the latest $125~{\rm GeV}$ Higgs data on the explanations.  

\subsection{Research strategy} \label{Section-analysis}
\begin{table}[tbp]
\caption{The parameter space explored in this study, assuming all the inputs are flatly distributed in prior since they have clear physical meanings. Considering that the soft trilinear coefficients for the third-generation squarks, $A_t$ and $A_b$, can significantly affect the SM-like Higgs boson mass by radiative corrections, we take $A_t = A_b$ and vary them. The unmentioned dimensional SUSY parameters are not crucial to this study, so we fix $M_3 =3~{\rm TeV}$, $m_A = 2~{\rm TeV}$, $m_C = 800~{\rm GeV}$, $v_s = 1~{\rm TeV}$, and a shared value of $2~{\rm TeV}$ for the others to be consistent with the LHC's search for new physics. We define all these parameters at the renormalization scale $Q_{input} = 1~{\rm TeV}$ and acquire the space by several trial scans over much broader parameter spaces.
\label{tab:scan}}
\centering

\vspace{0.3cm}

\resizebox{0.7\textwidth}{!}{
\begin{tabular}{c|c|c|c|c|c}
\hline
Parameter & Prior & Range & Parameter & Prior & Range   \\
\hline
$\lambda$ & Flat & $0.001 \sim 0.03$ & $\kappa$ & Flat & $-0.2 \sim 0.2$  \\
$\tan \beta$ & Flat & $5 \sim 60$ & $m_B/{\rm GeV}$ & Flat & $90 \sim 120$ \\
$\mu_{\rm tot}/{\rm TeV}$ & Flat & $0.4 \sim 1.0 $ &$m_N/{\rm TeV}$ & Flat & $-1.0 \sim 1.0$ \\
$A_t/{\rm TeV}$ & Flat & $1.0 \sim 3.0$ & $A_\lambda/{\rm TeV}$ & Flat & $ 1.0 \sim 3.0$ \\
$M_1/{\rm TeV}$ & Flat & $-1.0 \sim -0.2$ & $M_2/{\rm TeV}$ & Flat & $0.3 \sim 1.0$ \\
$M_{\tilde{\mu}_L}/{\rm TeV}$ & Flat & $0.2 \sim 1.0$ & $M_{\tilde{\mu}_R}/{\rm TeV}$ & Flat & $0.2 \sim 1.0$ \\
\hline
\end{tabular}}
\end{table}

\begin{table}[]
	\caption{Experimental analyses included in the package \texttt{SModelS-3.0.0}.}
	\label{tab:SModelS}
	\vspace{0.1cm}
	\resizebox{1.0 \textwidth}{!}{
		\begin{tabular}{cccc}
			\hline\hline
			\texttt{Name} & \texttt{Scenario} &\texttt{Final State} &$\texttt{Luminosity} (\texttt{fb}^{\texttt{-1}})$ \\\hline
   			\begin{tabular}[l]{@{}l@{}}ATLAS-1402-7029~\cite{Aad:1665444}\end{tabular}    & \begin{tabular}[c]{@{}c@{}c@{}c@{}} $\tilde{\chi}_1^{\pm}\tilde{\chi}_{2}^0\rightarrow \tilde{\ell}\nu\tilde{\ell}\ell(\tilde{\nu}\nu), \ell\tilde{\nu}\tilde{\ell}\ell(\tilde{\nu}\nu)\rightarrow \ell\nu\tilde{\chi}_1^0\ell\ell(\nu\nu)\tilde{\chi}_1^0$  \\ $\tilde{\chi}_1^{\pm}\tilde{\chi}_{2}^0\rightarrow\tilde\tau\nu\tilde\tau\tau(\tilde\nu\nu), \tau\tilde\nu\tilde\tau\tau(\tilde\nu\nu)\rightarrow\tau\nu\tilde{\chi}_1^0\tau\tau(\nu\nu)\tilde{\chi}_1^0$ \\ $\tilde{\chi}_1^{\pm}\tilde{\chi}_{2}^0\rightarrow W^\pm\tilde{\chi}_1^0Z\tilde{\chi}_1^0\rightarrow \ell\nu\tilde{\chi}_1^0\ell\ell\tilde{\chi}_1^0$ \\$\tilde{\chi}_1^{\pm}\tilde{\chi}_{2}^0\rightarrow W^\pm\tilde{\chi}_1^0h\tilde{\chi}_1^0\rightarrow \ell\nu\tilde{\chi}_1^0\ell\ell\tilde{\chi}_1^0$ \end{tabular}          & 3$ \ell$ + $ E_{\rm T}^{\rm miss}$    & 20.3 \\ \\
            \begin{tabular}[l]{@{}l@{}}CMS-SUS-16-034~\cite{CMS:2017kxn}\end{tabular}&$\tilde{\chi}_2^0\tilde{\chi}_1^{\pm}\rightarrow W\tilde{\chi}_1^0Z(h)\tilde{\chi}_1^0$ & n$\ell$(n\textgreater{}=2) + nj(n\textgreater{}=1) + $E_{\rm T}^{\rm miss}$       &               35.9               \\ \\
			\begin{tabular}[l]{@{}l@{}} CMS-SUS-16-039~\cite{CMS:2017moi} \end{tabular}          &\begin{tabular}[c]{@{}c@{}c@{}c@{}c@{}} $\tilde{\chi}_2^0\tilde{\chi}_1^{\pm}\rightarrow \ell\tilde{\nu}\ell\tilde{\ell}$\\$\tilde{\chi}_2^0\tilde{\chi}_1^{\pm}\rightarrow\tilde{\tau}\nu\tilde{\ell}\ell$\\$\tilde{\chi}_2^0\tilde{\chi}_1^{\pm}\rightarrow\tilde{\tau}\nu\tilde{\tau}\tau$\\ $\tilde{\chi}_2^0\tilde{\chi}_1^{\pm}\rightarrow WZ\tilde{\chi}_1^0\tilde{\chi}_1^0$\\$\tilde{\chi}_2^0\tilde{\chi}_1^{\pm}\rightarrow WH\tilde{\chi}_1^0\tilde{\chi}_1^0$\end{tabular} & n$\ell(n\textgreater{}0)$($\tau$) + $E_{\rm T}^{\rm miss}$& 35.9\\ \\			
         \begin{tabular}[l]{@{}l@{}}CMS-SUS-16-045~\cite{CMS:2017bki}\end{tabular}          &$\tilde{\chi}_2^0\tilde{\chi}_1^{\pm}\rightarrow W^{\pm}\tilde{\chi}_1^0h\tilde{\chi}_1^0$& 1$ \ell$  2b + $ E_{\rm T}^{\rm miss}$                           & 35.9               \\ \\
   			\begin{tabular}[l]{@{}l@{}}CMS-SUS-16-048~\cite{CMS:2018kag}\end{tabular}&\begin{tabular}[c]{@{}c@{}c@{}}$\tilde{\chi}_2^0\tilde{\chi}_1^{\pm}\rightarrow Z\tilde{\chi}_1^0W\tilde{\chi}_1^0$\\  $\tilde{\chi}_2^0\tilde{\chi}_1^0\rightarrow Z\tilde{\chi}_1^0\tilde{\chi}_1^0$\end{tabular}& 2$\ell$ + $E_{\rm T}^{\rm miss}$       &               35.9               \\ \\
			\begin{tabular}[l]{@{}l@{}} CMS-SUS-17-004~\cite{CMS:2018szt}\end{tabular} &$\tilde{\chi}_{2}^0\tilde{\chi}_1^{\pm}\rightarrow Wh(Z)\tilde{\chi}_1^0\tilde{\chi}_1^0$ & n$ \ell$(n\textgreater{}=0) + nj(n\textgreater{}=0) + $ E_{\rm T}^{\rm miss}$   & 35.9               \\ \\			\begin{tabular}[l]{@{}l@{}} CMS-SUS-17-009~\cite{CMS:2018eqb}\end{tabular}   &$\tilde{\ell}\tilde{\ell}\rightarrow \ell\tilde{\chi}_1^0\ell\tilde{\chi}_1^0$ &2$\ell$ + $E_{\rm T}^{\rm miss}$    &  35.9               \\ \\			\begin{tabular}[l]{@{}l@{}} CMS-SUS-17-010~\cite{CMS:2018xqw}\end{tabular}   &\begin{tabular}[c]{@{}c@{}}$\tilde{\chi}_1^{\pm}\tilde{\chi}_1^{\mp}\rightarrow W^{\pm}\tilde{\chi}_1^0 W^{\mp}\tilde{\chi}_1^0$\\$\tilde{\chi}_1^{\pm}\tilde{\chi}_1^{\mp}\rightarrow \nu\tilde{\ell} \ell\tilde{\nu}$ \\ \end{tabular}&2$ \ell$  + $E_{\rm T}^{\rm miss}$    & 35.9  \\ \\
   			\begin{tabular}[l]{@{}l@{}}ATLAS-1803-02762~\cite{ATLAS:2018ojr}\end{tabular} &\begin{tabular}[c]{@{}c@{}c@{}c@{}}$\tilde{\chi}_2^0\tilde{\chi}_1^{\pm}\rightarrow WZ\tilde{\chi}_1^0\tilde{\chi}_1^0$\\$\tilde{\chi}_2^0\tilde{\chi}_1^{\pm}\rightarrow \nu\tilde{\ell}l\tilde{\ell}$\\$\tilde{\chi}_1^{\pm}\tilde{\chi}_1^{\mp}\rightarrow \nu\tilde{\ell}\nu\tilde{\ell}$\\ $ \tilde{\ell}\tilde{\ell}\rightarrow \ell\tilde{\chi}_1^0\ell\tilde{\chi}_1^0$\end{tabular} & n$ \ell$ (n\textgreater{}=2) + $ E_{\rm T}^{\rm miss}$ & 36.1               \\ \\
   			\begin{tabular}[l]{@{}l@{}}ATLAS-1806-02293~\cite{ATLAS:2018eui}\end{tabular} &$\tilde{\chi}_2^0\tilde{\chi}_1^{\pm}\rightarrow WZ\tilde{\chi}_1^0\tilde{\chi}_1^0$ &n$\ell$(n\textgreater{}=2) + nj(n\textgreater{}=0) + $ E_T^{miss}$ & 36.1               \\ \\
			\begin{tabular}[l]{@{}l@{}}ATLAS-1812-09432~\cite{ATLAS:2018qmw}\end{tabular} &$\tilde{\chi}_2^0\tilde{\chi}_1^{\pm}\rightarrow Wh\tilde{\chi}_1^0\tilde{\chi}_1^0$ & n$ \ell$ (n\textgreater{}=0) + nj(n\textgreater{}=0) + nb(n\textgreater{}=0) + n$\gamma$(n\textgreater{}=0) + $E_{\rm T}^{\rm miss}$ & 36.1               \\ \\
            \begin{tabular}[l]{@{}l@{}}CMS-SUS-20-001~\cite{CMS:2020bfa}\end{tabular}&\begin{tabular}[c]{@{}c@{}c@{}}$\tilde{\chi}_2^0\tilde{\chi}_1^{\pm}\rightarrow Z\tilde{\chi}_1^0W\tilde{\chi}_1^0$\\ $\tilde{\ell}\tilde{\ell}\rightarrow \ell\tilde{\chi}_1^0\ell\tilde{\chi}_1^0$\end{tabular}& 2$\ell$ + $E_{\rm T}^{\rm miss}$       &               137               \\ \\
           \begin{tabular}[l]{@{}l@{}}CMS-SUS-20-004~\cite{Tumasyan:2799379}\end{tabular}&$\tilde{\chi}_2^0\tilde{\chi}_3^{0}\rightarrow h\tilde{\chi}_1^0h\tilde{\chi}_1^0$ & 4b +  $E_{\rm T}^{\rm miss}$       &               137               \\ \\
           \begin{tabular}[l]{@{}l@{}}CMS-SUS-21-002~\cite{Tumasyan:2809889}\end{tabular}&\begin{tabular}[c]{@{}c@{}c@{}c@{}}$\tilde{\chi}_1^{\pm}\tilde{\chi}_1^{\mp}\rightarrow W^{\pm}\tilde{\chi}_1^0 W^{\mp}\tilde{\chi}_1^0$ \\ $\tilde{\chi}_{2/3}^0\tilde{\chi}_1^{\pm}\rightarrow W^{\pm}\tilde{\chi}_1^0 Z(h)\tilde{\chi}_1^0$ \\ $\tilde{\chi}_2^0\tilde{\chi}_3^{0}\rightarrow Z\tilde{\chi}_1^0h\tilde{\chi}_1^0$ \end{tabular}& 4b + $E_{\rm T}^{\rm miss}$       &               137 \\ \\
			\begin{tabular}[l]{@{}l@{}}ATLAS-1908-08215~\cite{ATLAS:2019lff}\end{tabular}   &\begin{tabular}[c]{@{}c@{}}$\tilde{\ell}\tilde{\ell}\rightarrow \ell\tilde{\chi}_1^0\ell\tilde{\chi}_1^0$\\$\tilde{\chi}_1^{\pm}\tilde{\chi}_1^{\mp}$ \\ \end{tabular} & 2$\ell$ + $ E_{\rm T}^{\rm miss}$ & 139               \\ \\
			\begin{tabular}[l]{@{}l@{}}ATLAS-1909-09226~\cite{Aad:2019vvf}\end{tabular}          & $\tilde{\chi}_{2}^0\tilde{\chi}_1^{\pm}\rightarrow Wh\tilde{\chi}_1^0\tilde{\chi}_1^0$                            & 1$ \ell$ + h($\bm \rightarrow$ bb) + $ E_{\rm T}^{\rm miss}$    & 139 \\ \\
   			\begin{tabular}[l]{@{}l@{}}ATLAS-1911-12606~\cite{ATLAS:2019lng}\end{tabular}          & \begin{tabular}[c]{@{}c@{}c@{}c@{}} $\tilde{\chi}_{2}^0\tilde{\chi}_1^{\pm}\rightarrow Z\tilde{\chi}_1^0W\tilde{\chi}_1^0$ \\ $\tilde{\chi}_{2}^0\tilde{\chi}_1^{0}\rightarrow Z\tilde{\chi}_1^0\tilde{\chi}_1^0$ \\ $\tilde{\chi}_{1}^+\tilde{\chi}_1^-\rightarrow W^+\tilde{\chi}_1^0W^-\tilde{\chi}_1^0$ \\ $\tilde{\ell}\tilde{\ell}\rightarrow \ell\tilde{\chi}_1^0\ell\tilde{\chi}_1^0$\end{tabular}    & 2$ \ell$ + 2j + $ E_{\rm T}^{\rm miss}$    & 139 \\ \\
      		\begin{tabular}[l]{@{}l@{}}ATLAS-1912-08479~\cite{ATLAS:2019wgx}\end{tabular}          &$\tilde{\chi}_2^0\tilde{\chi}_1^{\pm}\rightarrow W(\rightarrow l\nu)\tilde{\chi}_1^0Z(\rightarrow\ell\ell)\tilde{\chi}_1^0$& 3$\ell $ + $ E_{\rm T}^{\rm miss}$                           & 139               \\ \\
           	\begin{tabular}[l]{@{}l@{}}ATLAS-2106-01676~\cite{ATLAS:2021moa}\end{tabular}          & \begin{tabular}[c]{@{}c@{}}$\tilde{\chi}_1^{\pm}\tilde{\chi}_{2}^0\rightarrow W(\rightarrow \ell \nu)Z(\rightarrow \ell\ell)\tilde{\chi}_1^0\tilde{\chi}_1^0$\\  $\tilde{\chi}_1^{\pm}\tilde{\chi}_{2}^0\rightarrow W(\rightarrow \ell \nu)h(\rightarrow \ell\ell)\tilde{\chi}_1^0\tilde{\chi}_1^0$                           \end{tabular} & 3$ \ell$ + $ E_{\rm T}^{\rm miss}$    & 139 \\ \\
   			\begin{tabular}[l]{@{}l@{}}ATLAS-2108-07586~\cite{ATLAS:2021yqv}\end{tabular}          & \begin{tabular}[c]{@{}c@{}}$\tilde{\chi}_1^{\pm}\tilde{\chi}_1^{\pm}\rightarrow WW\tilde{\chi}_1^0\tilde{\chi}_1^0$ \\ $\tilde{\chi}_1^{\pm}\tilde{\chi}_{2}^0\rightarrow WZ(h)\tilde{\chi}_1^0\tilde{\chi}_1^0$  \end{tabular}& 4j + $ E_{\rm T}^{\rm miss}$    & 139 \\ \\
      		\begin{tabular}[l]{@{}l@{}}ATLAS-2204-13072~\cite{Aad:2807817}\end{tabular}          & $\tilde{\chi}_{2}^0\tilde{\chi}_1^{\pm}\rightarrow W(\rightarrow q q
            )\tilde{\chi}_1^0Z(\rightarrow \ell\ell)\tilde{\chi}_1^0$                            & 2$ \ell$ + 2j + $ E_{\rm T}^{\rm miss}$    & 139 \\ \\
   			\begin{tabular}[l]{@{}l@{}}ATLAS-2209-13935~\cite{Aad:2834766}\end{tabular}          & \begin{tabular}[c]{@{}c@{}}$\tilde{\ell}\tilde{\ell}\rightarrow \ell\ell\tilde{\chi}_1^0\tilde{\chi}_1^0$ \\ $\tilde{\chi}_{1}^{\pm}\tilde{\chi}_1^{\mp}\rightarrow W(\rightarrow \ell \nu)W(\rightarrow \ell \nu)\tilde{\chi}_1^0\tilde{\chi}_1^0$     \end{tabular} & 2$ \ell$ + $E_{\rm T}^{\rm miss}$    & 139      \\ \hline\\
	\end{tabular}}
\end{table}

\begin{table}[]
	\caption{Experimental analyses of the electroweakino production processes considered in this study, which are categorized by the topologies of the SUSY signals.}
	\label{tab:LHC1}
	\vspace{0.2cm}
	\resizebox{0.97\textwidth}{!}{
		\begin{tabular}{llll}
			\hline\hline
			\texttt{Scenario} & \texttt{Final State} &\multicolumn{1}{c}{\texttt{Name}}\\\hline
			\multirow{6}{*}{$\tilde{\chi}_{2}^0\tilde{\chi}_1^{\pm}\rightarrow WZ\tilde{\chi}_1^0\tilde{\chi}_1^0$}&\multirow{6}{*}{$n\ell (n\geq2) + nj(n\geq0) + \text{E}_\text{T}^{\text{miss}}$}&\texttt{CMS-SUS-20-001($137fb^{-1}$)}~\cite{CMS:2020bfa}\\&&\texttt{ATLAS-2106-01676($139fb^{-1}$)}~\cite{ATLAS:2021moa}\\&&\texttt{CMS-SUS-17-004($35.9fb^{-1}$)}~\cite{CMS:2018szt}\\&&\texttt{CMS-SUS-16-039($35.9fb^{-1}$)}~\cite{CMS:2017moi}\\&&\texttt{ATLAS-1803-02762($36.1fb^{-1}$)}~\cite{ATLAS:2018ojr}\\&&\texttt{ATLAS-1806-02293($36.1fb^{-1}$)}~\cite{ATLAS:2018eui}\\\\
			\multirow{2}{*}{$\tilde{\chi}_2^0\tilde{\chi}_1^{\pm}\rightarrow \ell\tilde{\nu}\ell\tilde{\ell}$}&\multirow{2}{*}{$n\ell (n=3) + \text{E}_\text{T}^{\text{miss}}$}&\texttt{CMS-SUS-16-039($35.9fb^{-1}$)}~\cite{CMS:2017moi}\\&&\texttt{ATLAS-1803-02762($36.1fb^{-1}$)}~\cite{ATLAS:2018ojr}\\\\
			$\tilde{\chi}_2^0\tilde{\chi}_1^{\pm}\rightarrow \tilde{\tau}\nu\ell\tilde{\ell}$&$2\ell + 1\tau + \text{E}_\text{T}^{\text{miss}}$&\texttt{CMS-SUS-16-039($35.9fb^{-1}$)}~\cite{CMS:2017moi}\\\\
			$\tilde{\chi}_2^0\tilde{\chi}_1^{\pm}\rightarrow \tilde{\tau}\nu\tilde{\tau}\tau$&$3\tau + \text{E}_\text{T}^{\text{miss}}$&\texttt{CMS-SUS-16-039($35.9fb^{-1}$)}~\cite{CMS:2017moi}\\\\
			\multirow{6}{*}{$\tilde{\chi}_{2}^0\tilde{\chi}_1^{\pm}\rightarrow Wh\tilde{\chi}_1^0\tilde{\chi}_1^0$}&\multirow{6}{*}{$n\ell(n\geq1) + nb(n\geq0) + nj(n\geq0) + \text{E}_\text{T}^{\text{miss}}$}&\texttt{ATLAS-1909-09226($139fb^{-1}$)}~\cite{ATLAS:2020pgy}\\&&\texttt{CMS-SUS-17-004($35.9fb^{-1}$)}~\cite{CMS:2018szt}\\&&\texttt{CMS-SUS-16-039($35.9fb^{-1}$)}~\cite{CMS:2017moi}\\
			&&\texttt{ATLAS-1812-09432($36.1fb^{-1}$)}\cite{ATLAS:2018qmw}\\&&\texttt{CMS-SUS-16-034($35.9fb^{-1}$)}\cite{CMS:2017kxn}\\&&\texttt{CMS-SUS-16-045($35.9fb^{-1}$)}~\cite{CMS:2017bki}\\\\
			\multirow{2}{*}{$\tilde{\chi}_1^{\mp}\tilde{\chi}_1^{\pm}\rightarrow WW\tilde{\chi}_1^0 \tilde{\chi}_1^0$}&\multirow{2}{*}{$2\ell + \text{E}_\text{T}^{\text{miss}}$}&\texttt{ATLAS-1908-08215($139fb^{-1}$)}~\cite{ATLAS:2019lff}\\&&\texttt{CMS-SUS-17-010($35.9fb^{-1}$)}~\cite{CMS:2018xqw}\\\\
			\multirow{2}{*}{$\tilde{\chi}_1^{\mp}\tilde{\chi}_1^{\pm}\rightarrow 2\tilde{\ell}\nu(\tilde{\nu}\ell)$}&\multirow{2}{*}{$2\ell + \text{E}_\text{T}^{\text{miss}}$}&\texttt{ATLAS-1908-08215($139fb^{-1}$)}~\cite{ATLAS:2019lff}\\&&\texttt{CMS-SUS-17-010($35.9fb^{-1}$)}~\cite{CMS:2018xqw}\\\\
			\multirow{1}{*}{$\tilde{\chi}_2^{0}\tilde{\chi}_1^{\pm}\rightarrow ZW\tilde{\chi}_1^0\tilde{\chi}_1^0$}&\multirow{2}{*}{$2j(\text{large}) + \text{E}_\text{T}^{\text{miss}}$}&\multirow{2}{*}{\texttt{ATLAS-2108-07586($139fb^{-1}$)}~\cite{ATLAS:2021yqv}}\\{$\tilde{\chi}_1^{\pm}\tilde{\chi}_1^{\mp}\rightarrow WW\tilde{\chi}_1^0\tilde{\chi}_1^0$}&&\\\\
			\multirow{1}{*}{$\tilde{\chi}_2^{0}\tilde{\chi}_1^{\pm}\rightarrow (h/Z)W\tilde{\chi}_1^0\tilde{\chi}_1^0$}&\multirow{2}{*}{$j(\text{large}) + b(\text{large}) + \text{E}_\text{T}^{\text{miss}}$}&\multirow{2}{*}{\texttt{ATLAS-2108-07586($139fb^{-1}$)}~\cite{ATLAS:2021yqv}}\\{$\tilde{\chi}_2^{0}\tilde{\chi}_3^{0}\rightarrow (h/Z)Z\tilde{\chi}_1^0\tilde{\chi}_1^0$}&&\\\\
			$\tilde{\chi}_2^{0}\tilde{\chi}_1^{\mp}\rightarrow h/ZW\tilde{\chi}_1^0\tilde{\chi}_1^0,\tilde{\chi}_1^0\rightarrow \gamma/Z\tilde{G}$&\multirow{2}{*}{$2\gamma + n\ell(n\geq0) + nb(n\geq0) + nj(n\geq0) + \text{E}_\text{T}^{\text{miss}}$}&\multirow{2}{*}{\texttt{ATLAS-1802-03158($36.1fb^{-1}$)}~\cite{ATLAS:2018nud}}\\$\tilde{\chi}_1^{\pm}\tilde{\chi}_1^{\mp}\rightarrow WW\tilde{\chi}_1^0\tilde{\chi}_1^0,\tilde{\chi}_1^0\rightarrow \gamma/Z\tilde{G}$&&\\\\
			$\tilde{\chi}_2^{0}\tilde{\chi}_1^{\pm}\rightarrow ZW\tilde{\chi}_1^0\tilde{\chi}_1^0,\tilde{\chi}_1^0\rightarrow h/Z\tilde{G}$&\multirow{4}{*}{$n\ell(n\geq4) + \text{E}_\text{T}^{\text{miss}}$}&\multirow{4}{*}{\texttt{ATLAS-2103-11684($139fb^{-1}$)}~\cite{ATLAS:2021yyr}}\\$\tilde{\chi}_1^{\pm}\tilde{\chi}_1^{\mp}\rightarrow WW\tilde{\chi}_1^0\tilde{\chi}_1^0,\tilde{\chi}_1^0\rightarrow h/Z\tilde{G}$&&\\$\tilde{\chi}_2^{0}\tilde{\chi}_1^{0}\rightarrow Z\tilde{\chi}_1^0\tilde{\chi}_1^0,\tilde{\chi}_1^0\rightarrow h/Z\tilde{G}$&&\\$\tilde{\chi}_1^{\mp}\tilde{\chi}_1^{0}\rightarrow W\tilde{\chi}_1^0\tilde{\chi}_1^0,\tilde{\chi}_1^0\rightarrow h/Z\tilde{G}$&&\\\\
			\multirow{3}{*}{$\tilde{\chi}_{i}^{0,\pm}\tilde{\chi}_{j}^{0,\mp}\rightarrow \tilde{\chi}_1^0\tilde{\chi}_1^0+\chi_{soft}\rightarrow ZZ/H\tilde{G}\tilde{G}$}&\multirow{3}{*}{$n\ell(n\geq2) + nb(n\geq0) + nj(n\geq0) + \text{E}_\text{T}^{\text{miss}}$}&\texttt{CMS-SUS-16-039($35.9fb^{-1}$)}~\cite{CMS:2017moi}\\&&\texttt{CMS-SUS-17-004($35.9fb^{-1}$)}~\cite{CMS:2018szt}\\&&\texttt{CMS-SUS-20-001($137fb^{-1}$)}~\cite{CMS:2020bfa}\\\\
			\multirow{2}{*}{$\tilde{\chi}_{i}^{0,\pm}\tilde{\chi}_{j}^{0,\mp}\rightarrow \tilde{\chi}_1^0\tilde{\chi}_1^0+\chi_{soft}\rightarrow HH\tilde{G}\tilde{G}$}&\multirow{2}{*}{$n\ell(n\geq2) + nb(n\geq0) + nj(n\geq0) + \text{E}_\text{T}^{\text{miss}}$}&\texttt{CMS-SUS-16-039($35.9fb^{-1}$)}~\cite{CMS:2017moi}\\&&\texttt{CMS-SUS-17-004($35.9fb^{-1}$)}~\cite{CMS:2018szt}\\\\
			$\tilde{\chi}_{2}^{0}\tilde{\chi}_{1}^{\pm}\rightarrow W^{*}Z^{*}\tilde{\chi}_1^0\tilde{\chi}_1^0$&$3\ell + \text{E}_\text{T}^{\text{miss}}$&\texttt{ATLAS-2106-01676($139fb^{-1}$)}~\cite{ATLAS:2021moa}\\\\
			\multirow{3}{*}{$\tilde{\chi}_{2}^{0}\tilde{\chi}_{1}^{\pm}\rightarrow Z^{*}W^{*}\tilde{\chi}_1^0\tilde{\chi}_1^0$}&\multirow{2}{*}{$2\ell + nj(n\geq0) + \text{E}_\text{T}^{\text{miss}}$}&\texttt{ATLAS-1911-12606($139fb^{-1}$)}~\cite{ATLAS:2019lng}\\&&\texttt{ATLAS-1712-08119($36.1fb^{-1}$)}~\cite{ATLAS:2017vat}\\&&\texttt{CMS-SUS-16-048($35.9fb^{-1}$)}~\cite{CMS:2018kag}\\\\
			\multirow{3}{*}{$\tilde{\chi}_{2}^{0}\tilde{\chi}_{1}^{\pm}+\tilde{\chi}_{1}^{\pm}\tilde{\chi}_{1}^{\mp}+\tilde{\chi}_{1}^{\pm}\tilde{\chi}_{1}^{0}$}&\multirow{3}{*}{$2\ell + nj(n\geq0) + \text{E}_\text{T}^{\text{miss}}$}&\texttt{ATLAS-1911-12606($139fb^{-1}$)}~\cite{ATLAS:2019lng}\\&&\texttt{ATLAS-1712-08119($36.1fb^{-1}$)}~\cite{ATLAS:2017vat}\\&&\texttt{CMS-SUS-16-048($35.9fb^{-1}$)}~\cite{CMS:2018kag}\\\hline
	\end{tabular}}
\end{table}

\begin{table}[]
	\caption{Same as Table~\ref{tab:LHC1}, but for the slepton production processes.}
	\label{tab:LHC2}
  \centering
	\vspace{0.2cm}
	\resizebox{0.7\textwidth}{!}{
		\begin{tabular}{llll}
			\hline\hline
			\texttt{Scenario} & \texttt{Final State} &\multicolumn{1}{c}{\texttt{Name}}\\\hline
\multirow{6}{*}{$\tilde{\ell}\tilde{\ell}\rightarrow \ell\ell\tilde{\chi}_1^0\tilde{\chi}_1^0$}&\multirow{6}{*}{$2\ell + \text{E}_\text{T}^{\text{miss}}$}&\multirow{1}{*}{\texttt{ATLAS-1911-12606($139fb^{-1}$)}~\cite{ATLAS:2019lng}}\\&&\multirow{1}{*}{\texttt{ATLAS-1712-08119($36.1fb^{-1}$)}~\cite{ATLAS:2017vat}}\\&&\multirow{1}{*}{\texttt{ATLAS-1908-08215($139fb^{-1}$)}~\cite{ATLAS:2019lff}}\\&&\multirow{1}{*}{\texttt{CMS-SUS-20-001($137fb^{-1}$)}~\cite{CMS:2020bfa}}\\&&\multirow{1}{*}{\texttt{ATLAS-1803-02762($36.1fb^{-1}$)}~\cite{ATLAS:2018ojr}}\\&&\multirow{1}{*}{\texttt{CMS-SUS-17-009($35.9fb^{-1}$)}~\cite{CMS:2018eqb}}\\\hline
\end{tabular}} 
\end{table}

The discussion in Sec.~\ref{Section-Model}  shows that the three anomalies depend on the parameters $\lambda$, $\tan \beta$, $m_A$, $m_B$, $\mu_{\rm tot}$, $m_N$, $A_\lambda$, $M_1$, $M_2$, $M_{\tilde{\mu}_L}$, and $M_{\tilde{\mu}_R}$ in the GNMSSM framework.  Additionally, as demonstrated in Ref.~\cite{Meng:2024lmi}, the DM physics in the GNMSSM is also influenced by parameters $\kappa$, $A_\kappa$, $v_s$, and $m_C$. Since our primary objective is to 
investigate whether the GNMSSM can account for these three anomalies under various experimental constraints, rather than performing a comprehensive global fitting analysis of the model, we fix the value of some unimportant parameters in this study, including $m_A = 2~{\rm TeV}$, $m_C=800~{\rm GeV}$, and $v_s = 1~{\rm TeV}$. Furthermore, we determined $A_\kappa$ through the condition $\xi^\prime = 0 $, which yields~\cite{Meng:2024lmi}
 \begin{eqnarray}
 \kappa A_\kappa = \sqrt{2} m^2_B/v_s + \lambda \mu v^2/v_s^2 - \lambda(A_\lambda + m_N - \sqrt{2} \sin{2\beta} \kappa v_s) v^2/(2 v_s^2) + \sqrt{2} \kappa^2 v_s - 3 \kappa m_N, \nonumber 
 \end{eqnarray}
while allowing the remaining parameters to vary.

We performed a sophisticated scan over the parameter space delineated in Table~\ref{tab:scan}, using the MultiNest algorithm with ${\it{nlive}} = 12000$~\cite{MultiNest2009}\footnote{{\it{nlive}} in the MultiNest algorithm signifies the number of active or live points to determine the iso-likelihood contour in each iteration~\cite{MultiNest2009,Importance2019}.
 The larger it is, the more detailed the scan process will be in surveying the parameter space.}.
We constructed the following likelihood function to guide the scan:
\begin{eqnarray}
\mathcal{L} & \equiv&  \mathcal{L}_{\Delta a_\mu} \times \mathcal{L}_{\gamma \gamma + b \bar{b}} \times \mathcal{L}_{\rm Res} \\
\mathcal{L}_{\Delta a_\mu} &=& \exp \left[ -\frac{\chi^2_{\Delta a_\mu}}{2} \right] = \exp \left [ -\frac{1}{2} \left( \frac{a^{\rm SUSY}_\mu - 2.49\times 10^{-9}}{4.8\times 10^{-10}}\right)^2 \right]  \\
\mathcal{L}_{\gamma \gamma + b \bar{b}} &=& \exp \left[ -\frac{\chi^2_{\gamma \gamma + b\bar{b}}}{2} \right] =  \exp \left [-\frac{1}{2} \left( \frac{\mu_{\gamma\gamma} - 0.24}{0.08}\right)^2 -\frac{1}{2} \left( \frac{\mu_{b\bar{b}} - 0.117}{0.057}\right)^2 \right]_{m_{h_s} \simeq 95{\rm GeV}}.
\label{chi2-excesses}
\end{eqnarray}
Here, $\mathcal{L}_{\rm Res}$ represented the restrictions from relevant experiments on the theory: $\mathcal{L}_{\rm Res} = 1$ by our definition if the limitations were satisfied; otherwise, $\mathcal{L}_{\rm Res} = \exp\left [-100 \right ]$. These restrictions include

\begin{itemize}
\item \textbf{Masses of the light Higgs bosons:} Given the crucial importance of the radiative corrections to the SM-like Higgs boson mass in supersymmetric theories and the remarkable  precision of the LHC in determining its value,  the package \textsf{SARAH-4.15.3} incorporates all one-loop and 2L corrections when calculating the Higgs mass spectrum~\cite{Goodsell:2014bna,Goodsell:2015ira}. Specifically within the NMSSM framework, it has been recognized that the 2L correction to the mass beyond ${\cal{O}}(\alpha_s (\alpha_t +\alpha_b))$ - which constitutes the dominant 2L contribution in the MSSM - is also substantial, especially in scenarios that there is considerable mixing between the singlet and SM-doublet Higgs fields~\cite{Goodsell:2014pla}. This correction alone can shift $m_h$ by more than $2 {\rm~ GeV}$. Additionally, variations in the renormalization scale could modify the mass by over $2~{\rm GeV}$, even after incorporating the currently known corrections~\cite{Goodsell:2014pla}. In contrast, the radiation corrections to $m_{h_s}$ are less significant, as $h_s$ lacks potentially intense interactions with other fields. Considering that the mass of $h_s$ is estimated to be approximately $95.4~{\rm GeV}$ to account for the observed excesses and that of $h$ should be approximately $125~{\rm GeV}$ to align with the LHC measurements of the SM-like Higgs boson mass~\cite{ATLAS:2023oaq}, we set the theoretical and experimental uncertainties at $1~{\rm GeV}$ for $m_{h_s}$ and $3~{\rm GeV}$ for $m_h$. Consequently, this study establishes the mass ranges of $94.4~{\rm GeV} \leq m_{h_s} \leq 96.4~{\rm GeV}$ and $122~{\rm GeV} \leq m_{h} \leq 128~{\rm GeV}$. 
\item \textbf{Higgs data fit:} Given that $h$ corresponds to the discovered Higgs boson at the LHC, its properties should be consistent with corresponding measurements by the ATLAS and CMS collaborations at the $95\%$ confidence level. A p-value larger than 0.05 was essential, tested by the code \textsf{HiggsSignals-2.6.2}~\cite{HS2013xfa,HSConstraining2013hwa,HS2014ewa,HS2020uwn}.
\item \textbf{Extra Higgs searches:} This requirement is implemented through the code \textsf{HiggsBounds-5.10.2}~\cite{HB2008jh,HB2011sb,HBHS2012lvg,HB2013wla,HB2020pkv} and further via the code \textsf{HiggsTools-1.2}~\cite{Bahl:2022igd}.
\item \textbf{DM relic density:} We take the central value of $\Omega {h^2}=0.120$ from the Planck-2018 data~\cite{Planck:2018vyg} and assume theoretical uncertainties of $20\%$, i.e., $0.096 \leq \Omega {h^2} \leq 0.144$.
\item \textbf{DM detections:} The SI and SD DM-nucleon scattering cross-sections are required to be lower than corresponding bounds from the LZ experiments~\cite{LZ:2022ufs}. The DM indirect searches from the observation of dwarf galaxies by the Fermi-LAT collaboration are not considered as they impose no restrictions on the GNMSSM when $|m_{\tilde{\chi}_1^0}| \gtrsim 100~{\rm GeV}$~\cite{Fermi-LAT:2015att}.
\item \textbf{$B$-physics observables:} The branching ratios of $B_s \to \mu^+ \mu^-$ and $B \to X_s \gamma$ should be consistent with their experimental measurements at the $2\sigma$ level~\cite{pdg2018}.
\item \textbf{Vacuum stability:} The vacuum state of the scalar potential, composed of the Higgs fields and the last two-generation slepton fields\footnote{  
    In traditional investigations of vacuum stability constraints, researchers typically scrutinize the scalar potential encompassing the Higgs fields and the fields of the third-generation squarks and sleptons. Within the parameter space delineated in Table ~\ref{tab:scan}, the third-generation squark sector does not give rise to any charge and color breaking (CCB) phenomena, mainly because $A_t$ maintains a moderately small value. Specifically, it fulfills the CCB condition discussed in Ref.~\cite{Chowdhury:2013dka}:  $|A_t| \lesssim \sqrt{3(M_{\tilde{t}_L}^2 + M_{\tilde{t}_R}^2 + \mu_{tot}^2)}$. Moreover, with $M_{\tilde{\tau}_L} = M_{\tilde{\tau}_R} = 2~{\rm TeV}$, and $M_{\tilde{\mu}_L}$ and $M_{\tilde{\mu}_R}$ ranging within  $1~{\rm TeV}$, the second-generation slepton sector might impose stricter constraints on the theory in terms of vacuum stability compared to the third-generation slepton sector, as analyzed in Ref.~\cite{Cao:2021tuh}.},
should be either stable or long-lived, as indicated in Ref.~\cite{Hollik:2018wrr}.  In contrast with the MSSM, the GNMSSM exhibits enhanced vacuum stability due to its incorporation of the singlet Higgs field as a dynamical degree of freedom, particulary in scenarios with small $\lambda$ and large $\mu_{\rm tot}$ values, as elaborated in Ref.~\cite{Hollik:2018yek}. Nevertheless, excessively large soft-breaking trilinear coefficients may still lead to destabilization, as discussed in Ref.~\cite{Beuria:2016cdk}.  For our analysis, we utilized the package  \textsf{SARAH-4.15.3} to construct the model file for GNMSSM integration with \textsf{Vevacious} code ~\cite{Camargo-Molina:2013qva}, allowing both the slepton fields and the neutral components of the Higgs fields to acquire potential vevs. We evaluated the stability of this potential at zero temperature using the code \textsf{VevaciousPlusPlus}~\cite{VPP2014} (the C++ version of \textsf{Vevacious}).
\item \textbf{Unitarity constraint:} This constraint imposes significant requirements on the partial wave decomposition of $2 \to 2 $ scattering matrix elements, potentially limiting the maximum permissible values of the trilinear and quartic scalar couplings in the GNMSSM, as elaborated in Ref.~\cite{Goodsell:2018tti}. We implement it by the package \textsf{SARAH-4.15.3}.     
\end{itemize}

To get to know the impacts of the LHC's searches for supersymmetry on the scan results, the following processes were studied by concrete Monte Carlo simulations:
\begin{equation}\begin{split}
pp \to \tilde{\chi}_i^0\tilde{\chi}_j^{\pm} &, \quad i = 2, 3, 4, 5; \quad j = 1, 2 \\
pp \to \tilde{\chi}_i^{\pm}\tilde{\chi}_j^{\mp} &, \quad i,j = 1, 2; \\
pp \to \tilde{\chi}_i^{0}\tilde{\chi}_j^{0} &, \quad i,j = 2, 3, 4, 5; \\
pp \to \tilde{\mu}_i \tilde{\mu}_j^\ast &,\quad i,j = L, R; \\
pp \to \tilde{\nu}_\mu \tilde{\nu}_\mu^\ast.
\end{split}\end{equation}
Specifically, in order to save computing time, program \texttt{SModelS-3.0.0}~\cite{Khosa:2020zar}, which contains the selection efficiencies of the experimental analyses in Table~\ref{tab:SModelS},  was firstly used to exclude the obtained samples. Given that this program's capability in implementing the LHC constraints was limited by its database and the strict prerequisites to use it, the rest samples were further surveyed by simulating the analyses listed in Table~\ref{tab:LHC1} and ~\ref{tab:LHC2}. In this study, the cross-sections for each process were calculated to the next-to leading order by program \texttt{Prospino2}~\cite{Beenakker:1996ed}. 50000 events were generated for both electroweakino and slepton production processes, by package \texttt{MadGraph\_aMC@NLO}~\cite{Alwall:2011uj, Conte:2012fm}, and their parton shower and hadronization were finished by program \texttt{PYTHIA8}~\cite{Sjostrand:2014zea}. Detector simulations were implemented with program \texttt{Delphes}~\cite{deFavereau:2013fsa}. Finally, the event files were put into the package \texttt{CheckMATE\-2.0.26}~\cite{Drees:2013wra,Dercks:2016npn, Kim:2015wza} to calculate the $R$ value defined by $R \equiv max\{S_i/S_{i,obs}^{95}\}$ for all the involved analyses, where $S_i$ represents the simulated event number of the $i$-th signal region (SR), and $S_{i,obs}^{95}$ is
the corresponding $95\%$ confidence level upper limit. Evidently, $R > 1 $ indicates that the sample is experimentally excluded if the involved uncertainties are neglected~\cite{Cao:2021tuh}, while $R < 1$ means that it is consistent with the experimental analyses.

\begin{figure}[t]
\centering
\resizebox{1.02 \textwidth}{!}{
 \includegraphics{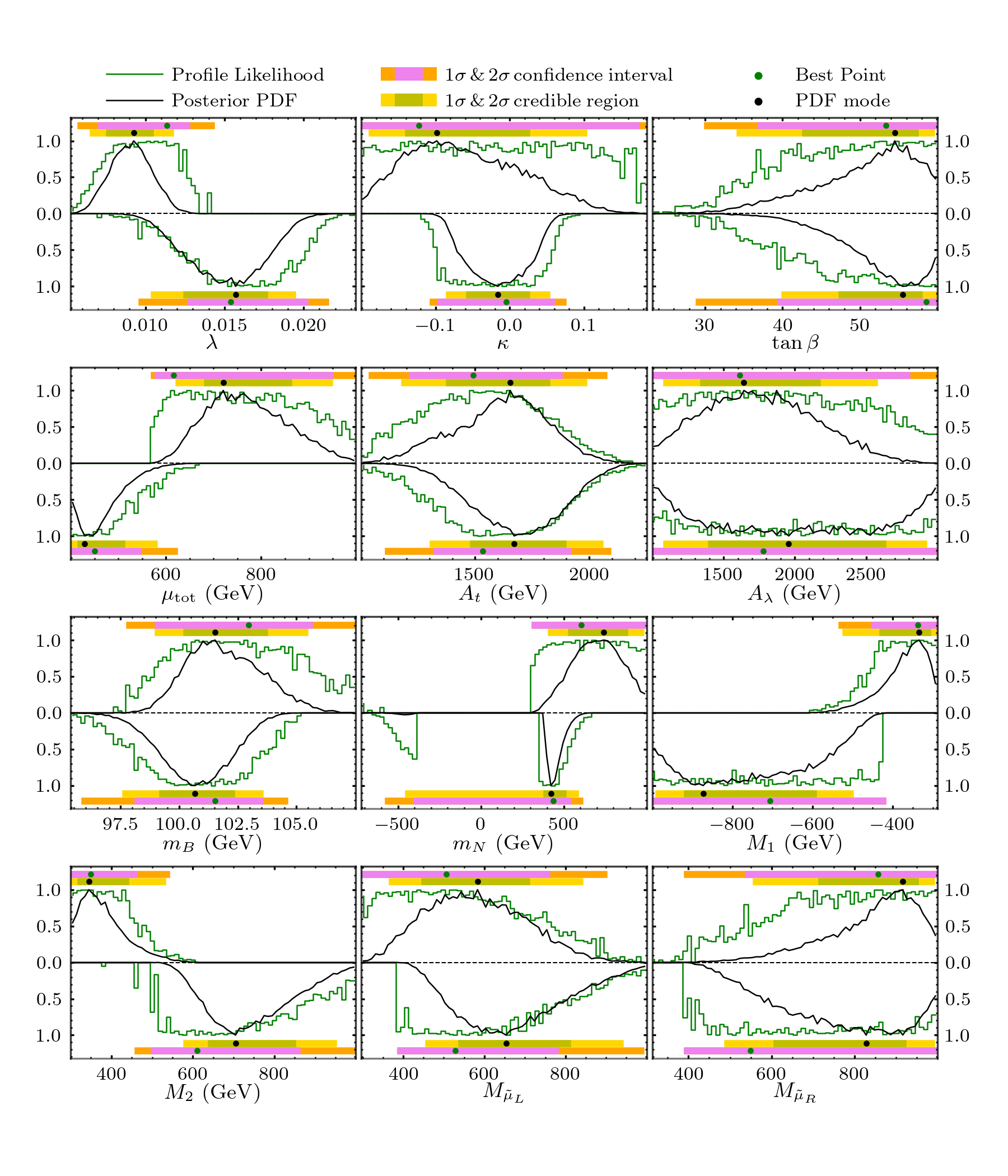}}
\vspace{-1.2cm}
\caption{One-dimensional profile likelihoods (green line) and posterior probability density functions (black line) of the input parameters $\lambda$, $\kappa$, $\tan \beta$, $\mu_{\rm tot}$, $A_t$, $A_\lambda$, $m_B$, $m_N$, $M_1$, $M_2$, $M_{\tilde{\mu}_L}$ and $M_{\tilde{\mu}_R}$ before implementing the restrictions from the LHC's search for supersymmetry. The upper (lower) part in each panel is obtained by the samples predicting the Bino-dominated (Singlino-dominated) DM. The violet and orange bands show the $1\sigma$ and $2\sigma$ confidence intervals, respectively, and the green dots mark the best point, corresponding to $\chi^2_{\gamma \gamma + b\bar{b}}+\chi^2_{\Delta a_\mu} =0.16$ for the Bino-DM case and $\chi^2_{\gamma \gamma + b\bar{b}}+\chi^2_{\Delta a_\mu} =0.02$ for the Singlino-DM case. The yellow and golden bands denote the $1\sigma$ and $2\sigma$ credible regions and the black dots are the modes of the posterior probability density function. All the statistical measures have been briefly introduced in Ref.~\cite{Fowlie:2016hew}. \label{Fig1}}
\end{figure}

\subsection{Numerical Results}

The scan process yields more than 92 thousand samples that are consistent with the experimental restrictions, among which about 50 thousand samples can simultaneously explain the muon $(g-2)$ anomaly and the diphoton and $b\bar{b}$ excesses at a level of 2$\sigma$. According to their predictions on the major component of DM, the obtained samples can be classified into the Bino-dominated case and the Singlino-dominated case. They achieve the measured relic abundance primarily by co-annihilating with the Wino-like and  Higgsino-like electroweakinos, respectively, and account for $26\%$ and $74\%$ of the total Bayesian evidence.

For both cases, we initially present in Fig.~\ref{Fig1} the statistical distributions of the twelve input parameters listed in Table~\ref{tab:scan}, including one-dimensional profile likelihoods (PL, delineated as green lines) and marginal posterior probability density functions (PDF, plotted as black lines)
\footnote{The profile likelihoods is the most significant likelihood value in a specific parameter space~\cite{Fowlie:2016hew}. Given a set of input parameters $\Theta \equiv (\Theta_1,\Theta_2,\cdots)$, one can acquire the one-dimensional PL by changing the other parameters to maximize the likelihood function, i.e.,
	\begin{eqnarray}
\mathcal{L}(\Theta_A)=\mathop{\max}_{\Theta_1,\cdots,\Theta_{A-1},\Theta_{A+1},\cdots}\mathcal{L}(\Theta), \nonumber
	\end{eqnarray}
The PL reflects the preference of a theory on the parameter space. For a given point $\Theta_A$, it represents the capability of the point in the theory to account for experimental data. By contrast, the one-dimensional marginal posterior probability density functions (PDF) of parameter $\Theta_A$  is obtained by integrating the posterior PDF from the Bayesian theorem, $P(\Theta)$, over the rest of the model inputs:
\begin{eqnarray}
	P(\Theta_A)&=&\int{P(\Theta) d\Theta_1 d\Theta_2 \cdots d\Theta_{A-1} d\Theta_{A+1} \cdots  \cdots }.  \nonumber
\end{eqnarray}
It reflects the preference for the samples acquired in the scan. }.
The distributions of the Bino-DM and Singlino-DM cases are shown in upper and lower parts in each panel, respectively. The violet and orange bands represent the $1\sigma$ and $2\sigma$ confidence intervals, and the green dots mark the best points corresponding to $\chi^2_{\gamma \gamma + b\bar{b}}+\chi^2_{\Delta a_\mu} =0.20$ for the Bino-DM case and $\chi^2_{\gamma \gamma + b\bar{b}}+\chi^2_{\Delta a_\mu} =0.04$ for the Singlino-DM case. The yellow and golden bands denote the $1\sigma$ and $2\sigma$ credible regions and the black dots denote the modes of the posterior probability density function.
The PL distributions indicate that the GNMSSM can account for the muon $(g-2)$ anomaly and the diphoton and $b\bar{b}$ excesses at a level of $2\sigma$ in broad parameter space except that $m_B$ is restricted within a narrow range. In fact, as noted in the approximations in Eqs.~(\ref{Approximation-relations}), $m_B$ is fixed by the relation $m_B^2 \simeq  m_{h_s}^2 |V_{h_s}^S|^2 + m_h^2 |V_{h_s}^{\rm SM}|^2$, where $V_{h_s}^{\rm S}$ and $V_{h_s}^{\rm SM}$ are determined by the signal rates of the excesses.

The distributions in Fig.~\ref{Fig1} reflect the impacts of the inputs on the observables involved in this study. They include
\begin{itemize}
\item $\mathbf{\mu_{tot}}$

In the Bino-DM case, the SI DM-nucleon scattering cross-section is inversely proportional to $\mu_{tot}^{2}$, with the LZ experiment alone stipulating a minimal $\mu_{tot}$ of approximately $380~{\rm GeV}$~\cite{Cao:2019qng}. When taking into account constraints from both the muon g-2 anomaly and results from the LHC, this minimum threshold increases to around 500 GeV~\cite{He:2023lgi}. Additionally, only explaining the diphoton and $b \bar{b}$ excesses at the $2\sigma$ level has set a lower bound on $\mu_{tot}$ of approximately $540~{\rm GeV}$~\cite{Cao:2023gkc}. Contrastingly, in the Singlino-DM scenario, the scattering rate notably diverges from the Bino-DM case as delineated by Eq. (2.30) in Ref.~\cite{Cao:2021ljw}, permitting a significantly lower $\mu_{tot}$. Specifically, as demonstrated in Ref.~\cite{Cao:2022ovk}, a $\mu_{tot}$ value as low as approximately 200 GeV is still viable even after incorporating the muon g-2 anomaly and LHC constraints. Such a $\mu_{tot}$ in the Singlino-DM case is also capable of explaining the diphoton and $b\bar{b}$ excesses~\cite{Cao:2023gkc}.

Furthermore, the ``WHL" contribution predominantly influences $a_\mu^{\rm SUSY}$, manifesting a declining trend as $\mu_{tot}$ enhances. This aspect elucidates the descending tendencies of the PL lines for $\mu_{tot}$ beyond approximately 750 GeV in Bino-DM and 425 GeV in Singlino-DM scenarios. Correspondingly, the posterior PDF curves of $\mu_{tot}$ peak around these values, highlighting the critical interplay between this parameter and experimental constraints.

After considering these factors, the comprehensive interpretation of the muon g-2 anomaly, as well as the diphoton and $b \bar{b}$ excesses, suggests a $2\sigma$ confidence interval for the $\mu_{tot}$ parameter within the scanning range presented in Table~\ref{tab:scan}: (580~{\rm GeV}, 1000~{\rm GeV}) for the Bino-DM case and (400~{\rm GeV}, 610 ~{\rm GeV}) for the Singlino-DM case.

\item $\mathbf{\lambda}$, $\mathbf{A_\lambda}$, and $\mathbf{m_N}$

Given the characteristics of $\mu_{tot}$ in both cases and the relation in Eq.~(\ref{Approximation-relations-1}), the Bino-DM case demonstrates a preference for a significantly smaller value of $\lambda$ compared to the Singlino-DM case when an appropriate $V_{h_s}^{\rm SM}$ is required to account for the observed excesses. Moreover, Eq.~(\ref{Approximation-relations-2}) suggests that the sum of $A_\lambda$ and $m_N$ should exhibit a much larger magnitude in the Bino-DM scenario than in the Singlino-DM scenario. To achieve this outcome, it is strongly favored for $m_N$ in the Bino-DM case to maintain a positive sign similar to $A_\lambda$ and vary within a wide range from $300~{\rm GeV}$ to $1000~{\rm GeV}$, where the lower bound arises from requiring $m_N > | M_1|$ to predict a Bino-DM, while the upper bound is defined by the constraints of the scanning area. By contrast, within the Singlino-DM framework, $|m_N| \simeq |m_{\tilde{\chi}_1^0}| \simeq \mu_{tot}$ is predicted since the DM primarily co-annihilates with the Higgsinos to obtain its measured abundance. Considering that explaining the muon g-2 anomaly at a $2\sigma$ level requires $m_{\tilde{\chi}_1^0} \lesssim 600~{\rm GeV}$~\cite{Cao:2022ovk} and $\mu_{tot}$ values listed in Table~\ref{tab:scan} are greater than 400~{\rm GeV}, one can deduce that $400~{\rm GeV} \lesssim |m_N| \lesssim 600~{\rm GeV}$. In addition, due to the cancellation between $m_N$ and $A_\lambda$ in Eq.~(\ref{Approximation-relations-2}), it becomes challenging for negative values of $m_N$ within the Singlino-DM case to account for the diphoton and $b\bar{b}$ excesses at a $1\sigma$ level. This feature leads to the suppression of the posterior PDF for a negative $m_N$, which is explicitly shown in Fig.~\ref{Fig1}.

It is noteworthy that in the Bino-DM scenario, the contribution of charginos to  $C_{h_s \gamma \gamma}$ is consistently less than $1\%$ of the SM contribution, attributed to the small values of $\lambda$ and the massiveness of the Higgsinos~\cite{Choi:2012he}. In contrast, as highlighted in our previous research~\cite{Cao:2023gkc}, this contribution can approach $1\%$ within the Singlino-DM framework. Additionally, the presence of a relatively substantial $\lambda$ value within the Singlino-DM framework proves advantageous for achieving the central values of $\mu_{\gamma \gamma}$ and $\mu_{b\bar{b}}$~\cite{Cao:2023gkc}. Collectively, these characteristics render the Singlino-DM scenario more suitable for explaining these excesses.

\item $\mathbf{M_1}$ and $\mathbf{M_2}$

In the GNMSSM, the impact of parameters $M_1$ and $M_2$ on the diphoton and $b\bar{b}$ signals is negligible. Therefore, their distributions are primarily determined by DM physics and the muon g-2 anomaly. In the case of Bino-DM, it is expected that the relation $|M_1| \simeq |m_{\tilde{\chi}_1^0}| \simeq M_2$ holds since the DM mainly co-annihilates with Wino-like electroweakinos to achieve the observed abundance. The permissible range for $|M_1|$ is approximately $210~{\rm GeV}$ to $620~{\rm GeV}$, where the lower threshold arises from constraints imposed by LHC in searching for sparticles, while the upper bound is derived from addressing the muon g-2 anomaly~\cite{He:2023lgi}. As $|M_1|$ exceeds around $450~{\rm GeV}$, it becomes challenging to effectively account for the muon g-2 anomaly at a $1\sigma$ level, resulting in diminished PL and posterior PDF distributions.
It is remarkable that a negative value of $M_1$ is preferred in this scenario as it can suppress SI DM-nucleon scattering through cancellation of different contributions~\cite{Cao:2019qng}. By contrast, the Singlino-DM scenario reveals distinct characteristics: $|M_1|$ only needs to surpass $|m_N|$ to predict appropriate dark matter properties, given that its influence on $\delta a_\mu^{\rm SUSY}$ is minor, and an effective explanation of muon g-2 anomaly requires $M_2 \lesssim 1.5~{\rm TeV}$~\cite{Cao:2022ovk}. Furthermore, when $M_2$ increases beyond $700~{\rm GeV}$, this scenario becomes challenging in explaining the muon g -2 anomaly at a $1\sigma$ level as reflected by diminishing PL and posterior PDF distributions of $M_2$.

\item $\mathbf{\tan \beta}$

As indicated by the condition to obtain a substantial $V_{h_s}^{\rm NSM} \tan \beta$ in Eq.~(\ref{Approximation-relations-2}) and the formulae of $\delta a_\mu^{\rm SUSY}$ in Eqs.~(\ref{eq:WHL}-\ref{eq:BLR}), a significant $\tan \beta$ is preferred for explaining both the muon g-2 anomaly and the diphoton/$b\bar{b}$ excesses. This characteristic is reflected by corresponding PL and PDF distributions in Fig.~\ref{Fig1}.

\item $\mathbf{M_{\tilde{\mu}_L}}$ and $\mathbf{M_{\tilde{\mu}_R}}$

In both scenarios, the impact of these two parameters is confined solely to $\delta a_\mu^{\rm SUSY}$. Notably, $\delta a_\mu^{\rm SUSY}$ exhibits a greater sensitivity to $M_{\tilde{\mu}_L}$ in comparison to $M_{\tilde{\mu}_R}$, leading to a preference of a moderately light $M_{\tilde{\mu}_L}$ when addressing the muon g-2 anomaly. In addition to the PL and PDF plots of these parameters shown in Fig.~\ref{Fig1}, we will delve deeper into the smuon spectrum through scattering plots in subsequent discussions.

\item $\mathbf{A_t}$

The trilinear coefficient $A_t$ significantly influences the SM-like Higgs boson mass $m_h$, the mixing $V_{h_s}^{\rm SM}$, and the $h_s g g$ and $h_s \gamma \gamma$ coupling strengths via Stop-mediated loops. Given that we have fixed the soft-breaking masses of squarks at $2~{\rm TeV}$, $A_t$s around $1.7~{\rm TeV}$ are preferred to make the theory accessible to explain the excesses.

\item $\mathbf{\kappa}$

Regarding the Bino-DM case, Eqs.~(\ref{Simplify-1}) and (\ref{New-mass-matrix}) indicate that $\kappa$ does not directly impact the observables involved in this study, except for its influence on the scalar potential of the singlet field and consequently affecting the electroweak symmetry breaking. Therefore, the naturalness in realizing the symmetry breaking determines its posterior PDF~\cite{Cao:2018iyk}. In the framework of the Singlino-DM scenario, however, there is an additional effect related to $\kappa$: since the amplitude of the $t$-channel $h_s$-mediated contribution to the SI DM-nucleon scattering is proportional to $\kappa V_{h_s}^{\rm SM}$ in the leading-order approximation,
and considering this contribution's potential significance, the LZ results favor a small value of $\kappa$~\cite{Cao:2021ljw}.

\end{itemize}
We note that the PLs and posterior PDFs depicted in Fig.~\ref{Fig1} can significantly differ from those presented in Ref.~~\cite{Cao:2023gkc}, which were derived by a similar methodology but neglected the muon g-2 anomaly in the likelihood function.

\begin{figure}[t]
		\centering
\includegraphics[width=0.99\linewidth]{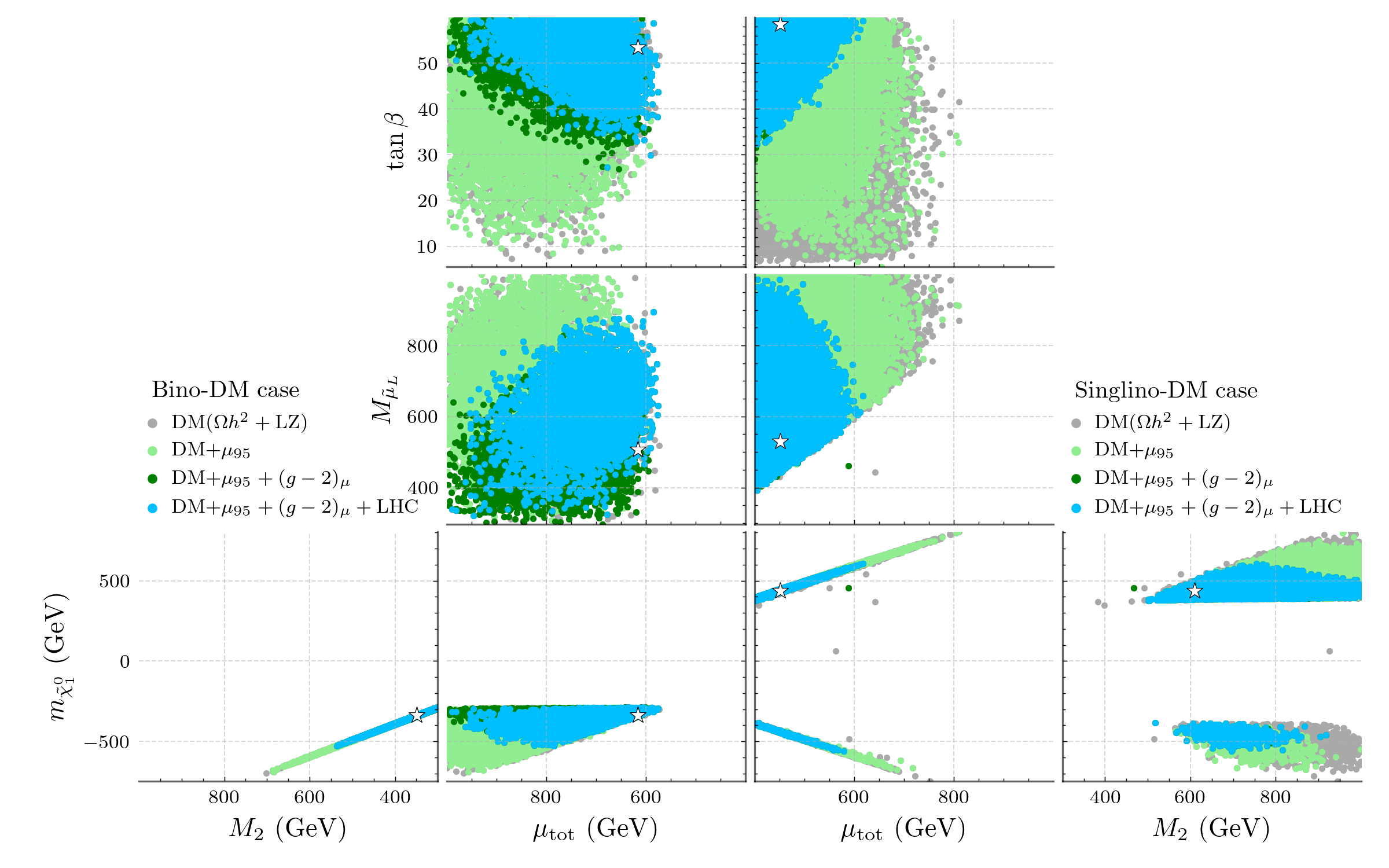}
\vspace{-0.5cm}
\caption{The samples for the Bino-DM case obtained by the scan in Sec. 3.1 and those for the Singlino-DM case, which are projected onto the $\mu_{\rm tot}-\tan\beta$, $\mu_{\rm tot}-M_{\tilde{\mu}_L}$, $\mu_{\rm tot}-m_{\tilde{\chi}_1^0}$,  and $m_{\tilde{\chi}_1^0} - M_2$ planes in the left and right sections, respectively.  The colors distinguish four combinations of constraints from the latest DM detections, the $\gamma\gamma$ and $b\bar{b}$ signal rates with $\chi^2_{\gamma \gamma + b\bar{b}} \leq 2.3$ ($1\sigma$ level for two degrees of freedom), the muon $g-2$ measurement with $\chi^2_{\Delta a_\mu} \leq 4.0$ ($2\sigma$ level), and the LHC's search for sparticles. \label{Fig2} }
\end{figure}

Secondly, we generate a bilateral diagram in which we project the samples of the Bino-DM case and those of the Singlino-DM case onto the $\mu_{\rm tot}-\tan\beta$, $\mu_{\rm tot}-M_{\tilde{\mu}_L}$, $\mu_{\rm tot}-m_{\tilde{\chi}_1^0}$, and $m_{\tilde{\chi}_1^0} - M_2$ planes in the left and right sections of Fig.~\ref{Fig2}, respectively.  The different colors distinguish four combinations of the constraints from the latest DM detections, the $\gamma\gamma$ and $b\bar{b}$ signal rates with $\chi^2_{\gamma \gamma + b\bar{b}} \leq 2.3$, the muon $g-2$ measurement with $\chi^2_{\Delta a_\mu} \leq 4.0$, and the LHC's search for sparticles. As a useful supplement to Fig.~\ref{Fig2}, we also list in Table~\ref{tab:numbers} the numbers of these colored samples. From these scattering plots and Table~\ref{tab:numbers}, one can infer the following facts:

\begin{table}[t]
\centering
\resizebox{1\textwidth}{!}
{
\begin{tabular}{l|c|c|c|c}
  \hline
  \hline
 \diagbox[height=2.7em]{Scenario}{Number}{Constraints}
 & \begin{tabular}[c]{@{}c@{}} DM$(\Omega h^2 + {\rm LZ})$ \\ Grey\end{tabular}
 & \begin{tabular}[c]{@{}c@{}} DM + $\mu_{\rm 95}$ \\ Green \end{tabular}
 &\begin{tabular}[c]{@{}c@{}}  DM + $\mu_{\rm 95}$+$(g-2)_\mu$ \\ Dark Green \end{tabular}
 &\begin{tabular}[c]{@{}c@{}}  DM +$\mu_{\rm 95}$ +$(g-2)_\mu$ + LHC \\ Blue \end{tabular} \\ \hline
  Bino-DM 						& 17368 & 14416 & 8408 & 4770 \\
  \hline
  Singlino-DM 					& 59038 & 38731 & 20851 & 20539 \\
  \hline
  \hline
\end{tabular}}
\caption{\label{tab:numbers} Numbers of the samples in the Bino- and Singlino-DM scenarios, respectively, classified by the experimental constraints they satisfy. In Fig.~\ref{Fig2}, these samples are labelled with grey, green, dark green, and blue colors, respectively.}
\end{table}

\begin{figure}[t]
	\centering
	\includegraphics[width=0.9\linewidth]{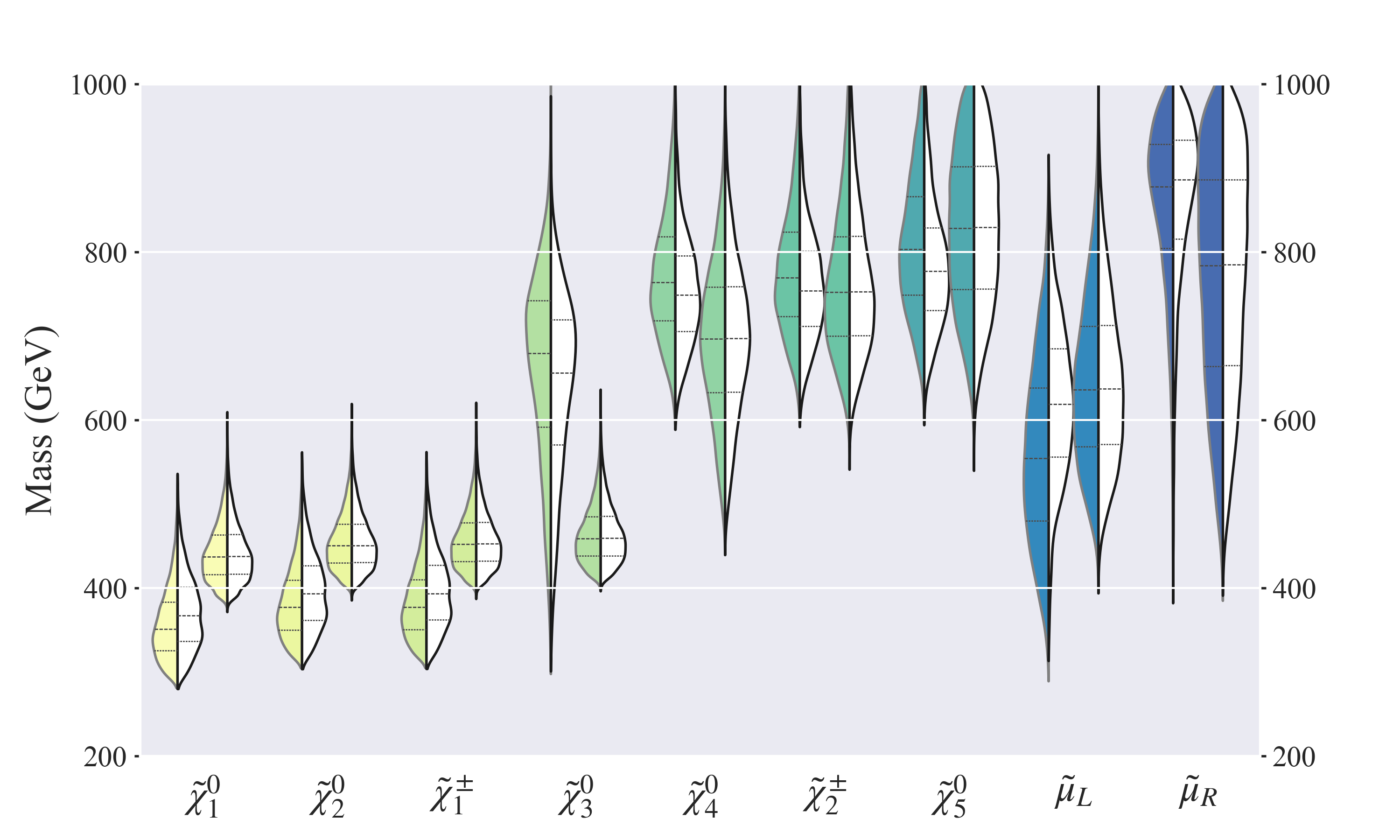}
\vspace{-0.3cm}
\caption{Mass distribution for the involved sparticles, which are scaled by counting sample numbers. The left and right violins for each sparticle represent the Bino- and Singlino-DM cases, respectively. The colored  half in each violin is plotted with the dark green samples in Fig.~\ref{Fig2}, while the white is drawn with the blue samples. The dashed line labels the quartile ranges. \label{Fig3}}
\end{figure}

\begin{itemize}
\item
The $m_{\tilde{\chi}_1^0}-M_2$ panel for the Bino-DM case demonstrates that the DM achieves the measured relic abundance through co-annihilating with the Wino-like electroweakinos. We verified that the dominant contribution to this abundance arises from the channels $\tilde{\chi}_2^0 \tilde{\chi}_1^- \to d_i \bar{u}_i$, where $i=1,2$ represents the first two generations of quarks. Additionally, our analysis reveals that the scenarios where Bino-like DM co-annihilates with Wino-like electroweakinos along with either muon-flavored sneutrino or smuons contribute less than 1\% of total Bayesian evidence.
\item
The $m_{\tilde{\chi}_1^0}-\mu_{tot}$ panel for the Singlino-DM case indicates that the DM predominantly co-annihilated with Higgsino-like electroweakinos to attain the measured relic abundance. Our analysis suggests that there is also a possibility of simultaneous co-annihilation of the DM with Higgsinos and muon-flavored sleptons; however, this scenario only contributes to less than 0.2\% of the total Bayesian evidence. Moreover, we have identified a few samples featuring Singlino-Wino co-annihilation. Nevertheless, these samples pose challenges in explaining the 95 GeV excesses at a $1\sigma$ level due to the reason listed in the following item.

\item The condition stated in Eq.~(\ref{Approximation-relations-1}) is crucial for the GNMSSM to address the diphoton and $b\bar{b}$ excesses, while its impact on forecasting the DM observables is negligible. This elucidates why some samples capable of interpreting the DM experiments fail to explain the excesses, as shown in Fig.~\ref{Fig2} and Table~\ref{tab:numbers}.

\item The impacts of $\tan \beta$, $M_1$, $M_2$, $\mu_{tot}$, $M_{\tilde{\mu}_L}$, and $M_{\tilde{\mu}_R}$ on $\delta a_\mu^{\rm SUSY}$ are delineated in Eqs.~~(\ref{eq:WHL}-\ref{eq:BLR}). Among these parameters, only $\mu_{tot}$ plays a crucial role in explaining the excesses at 95 GeV. This characteristic accounts for the distinct regions in each panel of Fig.~\ref{Fig2} favored by the excesses and the muon g-2 anomaly.

\item  The LHC search for sparticles has long been acknowledged as a stringent constraint on the SUSY explanations of the muon g-2 anomaly (see, for example, the analysis presented in Ref.~\cite{He:2023lgi}).  In relation to this study, the manifestation of this phenomenon is clearly observed in the Bino-DM scenario, while it remains ambiguous in the Singlino-DM scenario (refer to Table~\ref{tab:numbers}).  This difference is attributed to the substantial mass of the DM particle in the latter case (specifically, $|m_{\tilde{\chi}_1^0}| \gtrsim 400~{\rm GeV}$ in the Singlino-DM case versus $|m_{\tilde{\chi}_1^0}| \gtrsim 300~{\rm GeV}$ in the Bino-DM case), which in turn diminishes the restrictions imposed by the LHC.

Additionally, our analysis reveals that the maximal prediction of $\delta a_\mu^{\rm SUSY}$ diminishes as $|m_{\tilde{\chi}_1^0}|$ increases, once the constraints from the LHC are accounted for. In the Bino-DM scenario, $\delta a_\mu^{\rm SUSY}$ can reach a value of $2.7 \times 10^{-9}$ when $|m_{\tilde{\chi}_1^0}|$ is approximately $300~{\rm GeV}$. By contrast, in the Singlino-DM scenario, it can achieve a value of $3.0 \times 10^{-9}$ for $|m_{\tilde{\chi}_1^0}|$ around $400~{\rm GeV}$. The ability to predict a higher $\delta a_\mu^{\rm SUSY}$ at the same $|m_{\tilde{\chi}_1^0}|$ in the latter scenario arises from the more complex decay chains of the involved sparticles,
leading to less stringent LHC constraints. This aspect was emphasized in our prior work~\cite{Cao:2022ovk}.

\item
The unified explanation and the LHC restrictions set bounds on $\tan\beta$ and the masses of the involved sparticles, i.e., $\tan\beta \gtrsim 26$, $580 ~{\rm GeV} \lesssim \mu_{\rm tot} \lesssim  1000~{\rm GeV}$, $300 ~{\rm GeV} \lesssim |m_{\tilde{\chi}_1^0}|, M_2 \lesssim  550~{\rm GeV}$, and $300 ~{\rm GeV} \lesssim M_{\tilde{\mu}_L} \lesssim  900~{\rm GeV}$ for the Bino-DM case and $\tan\beta \gtrsim 32$, $400 ~{\rm GeV} \lesssim |m_{\tilde{\chi}_1^0}|, \mu_{\rm tot} \lesssim  600~{\rm GeV}$, $500 ~{\rm GeV} \lesssim M_2 \lesssim  1000~{\rm GeV}$ and $400 ~{\rm GeV} \lesssim M_{\tilde{\mu}_L} \lesssim  1000~{\rm GeV}$ for the Singlino-DM case. To display the bounds more clearly, we plot the mass distributions of the involved sparticles by `Violin' diagram in Fig.~\ref{Fig3}.
\end{itemize}

\begin{figure}[t]
	\centering
	\includegraphics[width=0.9\linewidth]{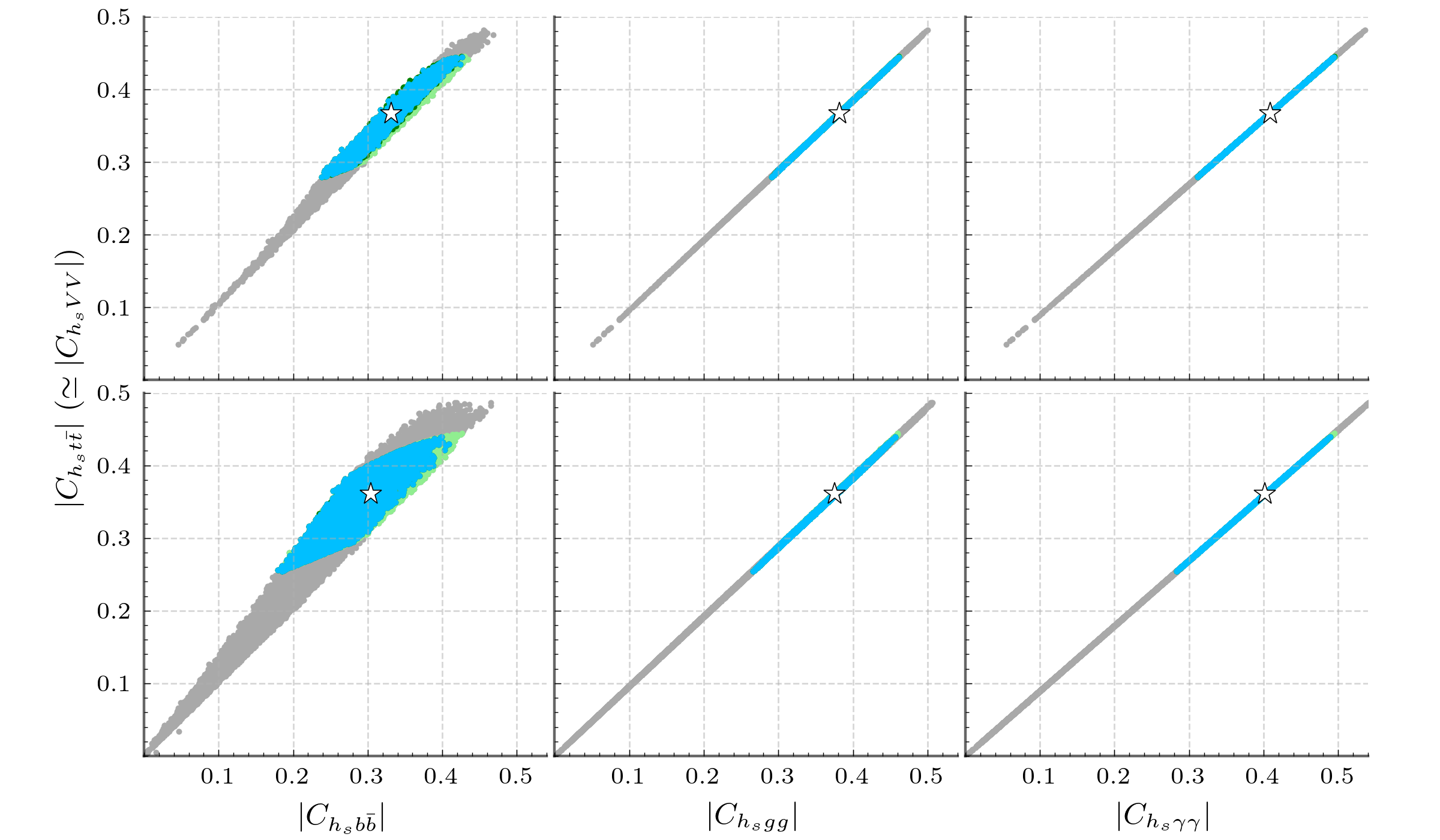}
\vspace{-0.3cm}
\caption{Projection of the samples in Fig.~\ref{Fig2} onto $|C_{h_s b \bar{b}}|-|C_{h_s t \bar{t}}|$, $|C_{h_s g g}|-|C_{h_s t \bar{t}}|$, and $|C_{h_s \gamma \gamma}|-|C_{h_s t  \bar{t}}|$ planes. The upper and lower panels correspond to the Bino- and Singlino-DM cases, respectively.  \label{Fig4}}
\end{figure}

\begin{figure}[t]
	\centering
	\includegraphics[width=0.9\linewidth]{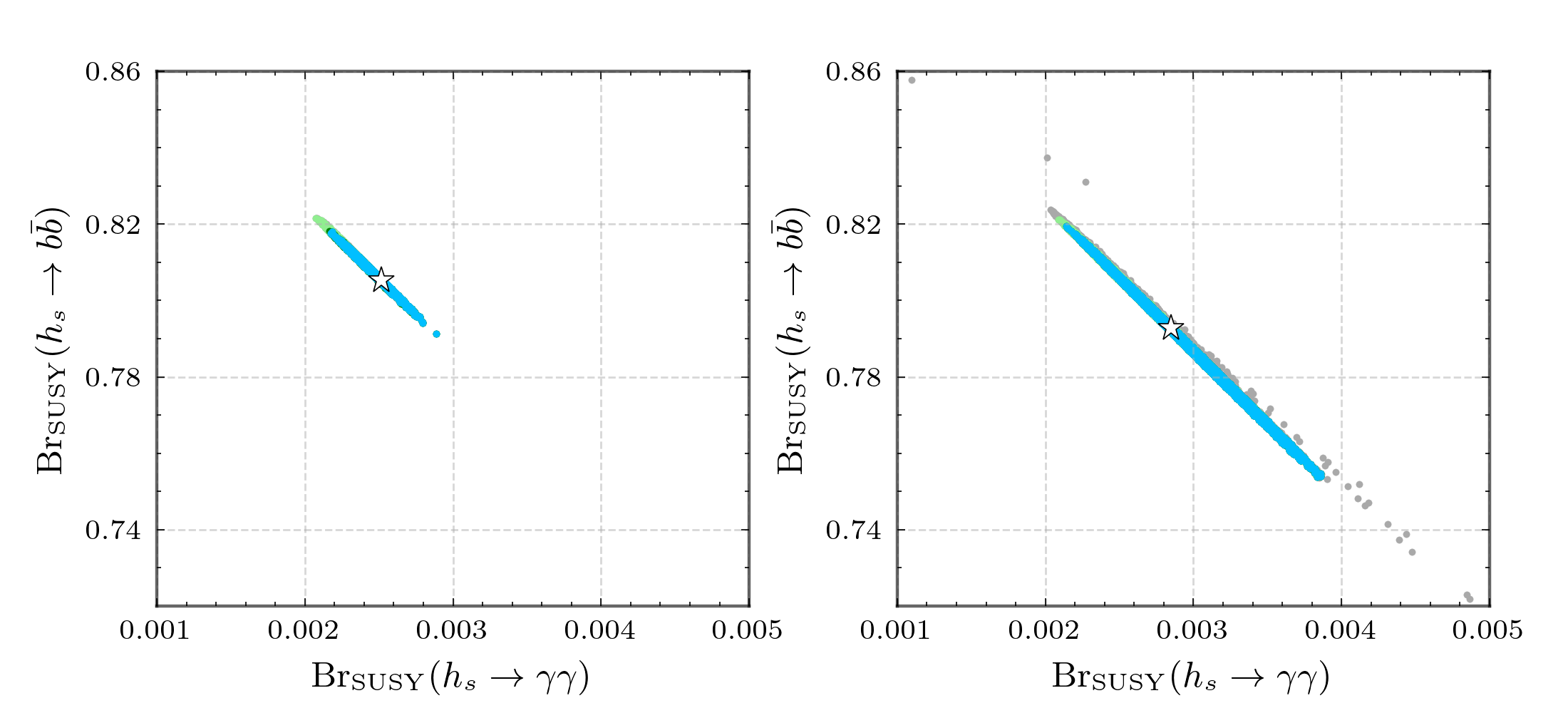}
\vspace{-0.3cm}
\caption{ Projection of the samples in Fig.~\ref{Fig2} onto ${\rm Br}_{\rm SUSY}(h_s \to \gamma \gamma) - {\rm Br}_{\rm SUSY}(h_s \to b \bar{b})$ planes. The left and right panels correspond to the Bino- and Singlino-DM cases, respectively.  \label{Fig5}}
\end{figure}

Thirdly, we study the properties of the Higgs bosons. We project the samples in Fig.~\ref{Fig2} onto $|C_{h_s b \bar{b}}|-|C_{h_s t \bar{t}}|$, $|C_{h_s g g}|-|C_{h_s t \bar{t}}|$, and $|C_{h_s \gamma \gamma}|-|C_{h_s t  \bar{t}}|$ planes to obtain Fig.~\ref{Fig4} and onto ${\rm Br}_{\rm SUSY}(h_s \to \gamma \gamma) - {\rm Br}_{\rm SUSY}(h_s \to b \bar{b})$ planes to yield Fig.~\ref{Fig5}. These figures reveal that both $|C_{h_s g g}|$ and $|C_{h_s \gamma \gamma}|$ are slightly higher than $|C_{h_s t \bar{t}}|$, while $|C_{h_s b \bar{b}}|$ is relatively suppressed. Additionally, there is a strong correlation between $Br_{\rm SUSY} (h_s \to \gamma \gamma)$ and $Br_{\rm SUSY} (h_s \to b \bar{b})$ in explaining the excesses at 95 GeV, whereas $C_{h_s b \bar{b}}$ loosely correlates with $C_{h_s t \bar{t}}$. The best point for the Bino-DM is located at $|C_{h_s t \bar{t}}| \simeq |C_{h_s V V}| \simeq 0.37$, $|C_{h_s b \bar{b}}| \simeq 0.33$, $|C_{h_s gg}| \simeq 0.38$, and $|C_{h_s \gamma\gamma}| \simeq 0.41$, leading to ${\rm Br}_{\rm SUSY} (h_s \to \gamma \gamma) \simeq 0.25 \%$, ${\rm Br}_{\rm SUSY} (h_s \to b \bar{b}) \simeq 81\%$, $\mu_{\gamma \gamma} = 0.22$, $\mu_{b\bar{b}} = 0.13$, and $\chi^2_{\gamma \gamma + b \bar{b}} = 0.12 $. In contrast, the optimal configuration for the Singlino-DM case corresponds to $|C_{h_s t \bar{t}}| \simeq |C_{h_s V V}| \simeq 0.35$, $|C_{h_s b \bar{b}}| \simeq 0.29$, $|C_{h_s gg}| \simeq 0.37$, and $|C_{h_s \gamma\gamma}| \simeq 0.39$, predicting ${\rm Br}_{\rm SUSY} (h_s \to \gamma \gamma) \simeq 0.28 \%$, ${\rm Br}_{\rm SUSY} (h_s \to b \bar{b}) \simeq 79\%$, $\mu_{\gamma \gamma} = 0.24$, $\mu_{b\bar{b}} = 0.12$, and $\chi^2_{\gamma \gamma + b \bar{b}} \simeq 0 $. It is noteworthy that the Singlino-DM case allows significantly wider ranges of the couplings to account for the excesses at the $1\sigma$ level when compared to the Bino-DM case, revealing its greater adaptability in forecasting Higgs observables. The underlying reasons for these phenomena was discussed in Sec. 2.3.

\begin{figure}[t]
	\centering
	\includegraphics[width=0.9\linewidth]{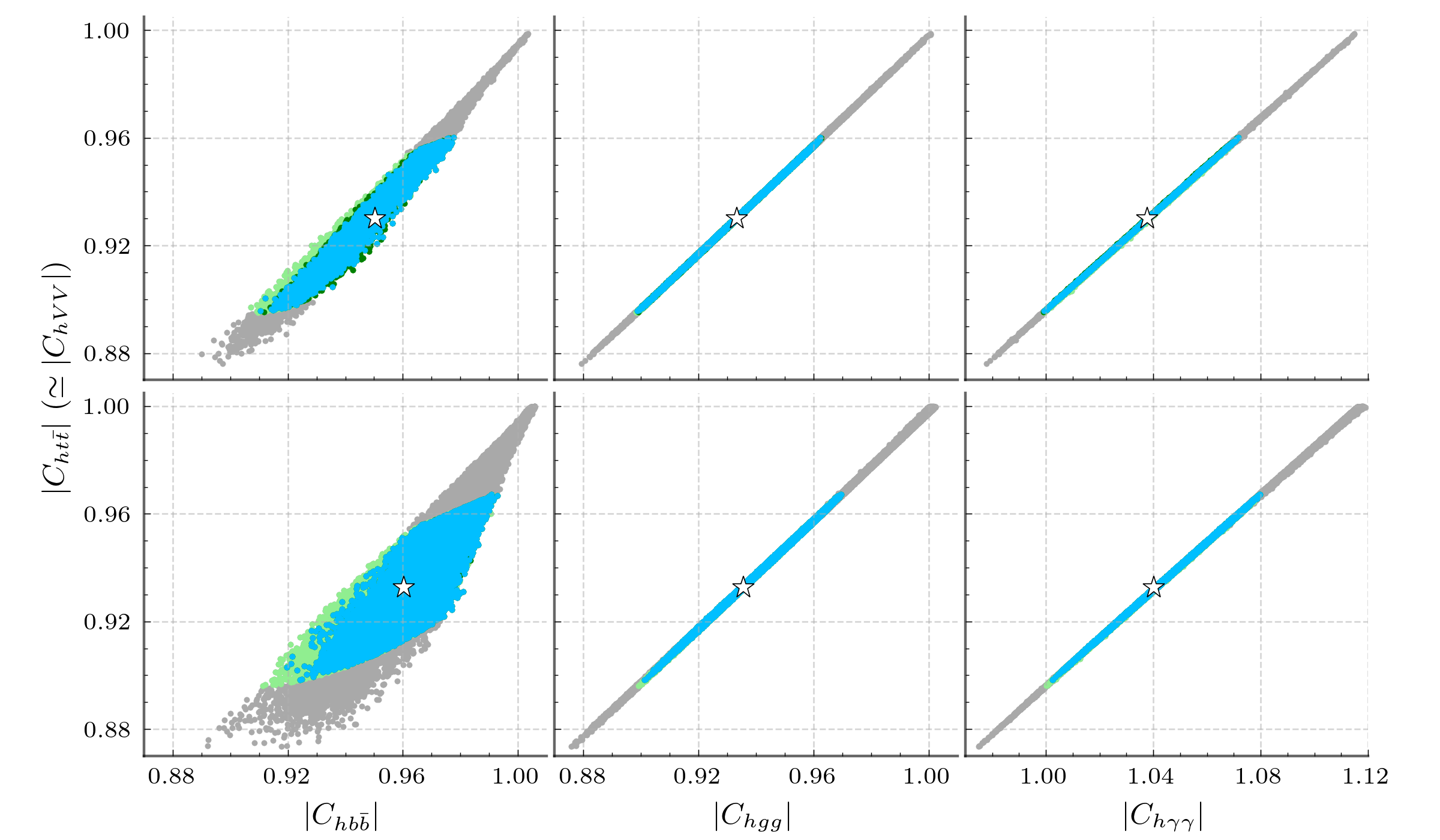}
\vspace{-0.3cm}
\caption{The same as Fig.~\ref{Fig4}, except that the couplings of the SM-like Higgs boson are displayed.   \label{Fig6}}
\end{figure}

\begin{figure}[t]
	\centering
	\includegraphics[width=0.9\linewidth]{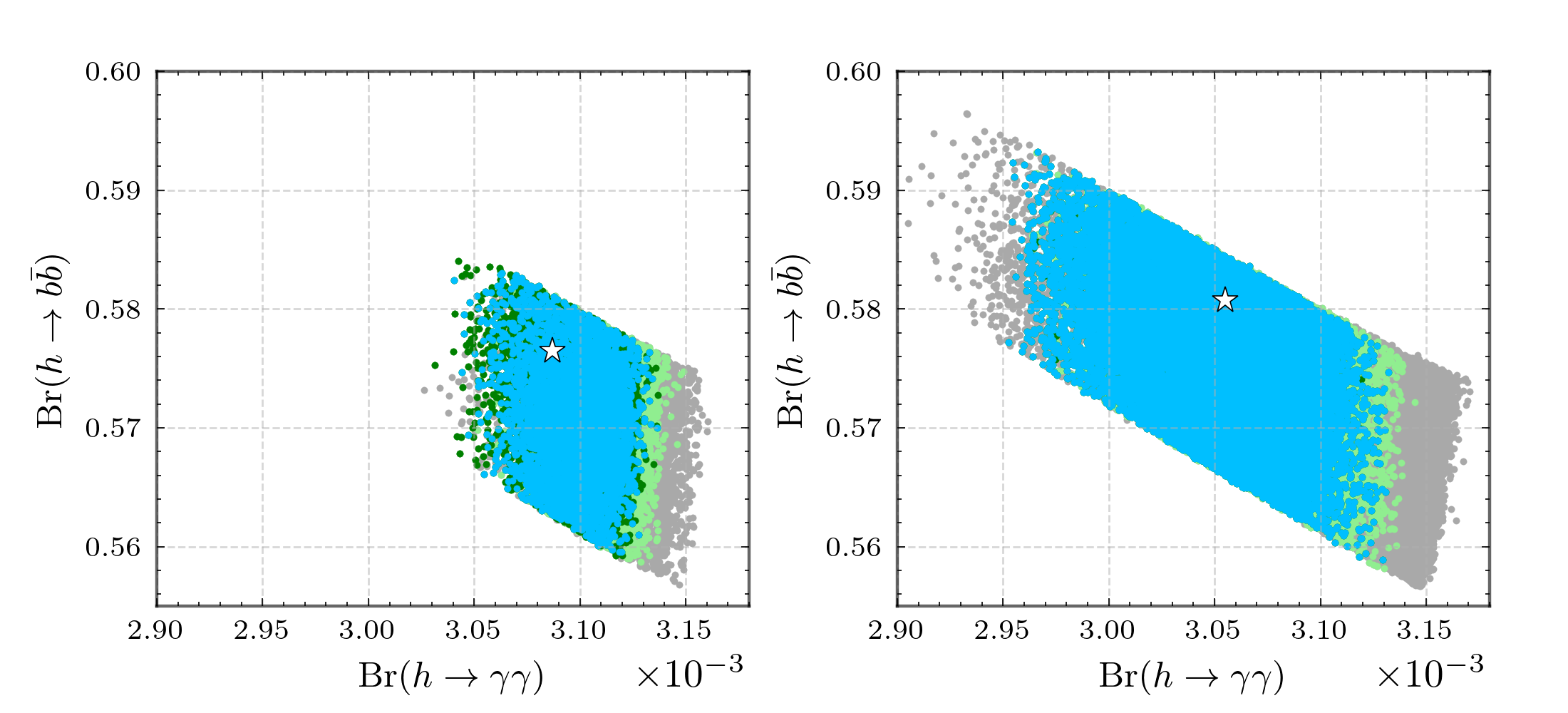}
\vspace{-0.3cm}
\caption{The same as Fig.~\ref{Fig5}, except that the branching ratios of the SM-like Higgs boson are shown.   \label{Fig7}}
\end{figure}

\begin{table}[t]
\centering
\resizebox{1\textwidth}{!}
 {
\begin{tabular}{lrlr|lrlr}
\hline \hline
\multicolumn{4}{c|}{\bf Benchmark Point P1}                                                                                                								& \multicolumn{4}{c}{\bf Benchmark Point P2}
 \\ \hline
$\lambda$             		& 0.011			& $A_t$                			& 1491~GeV		&$\lambda$             			& 0.015			& $A_t$                			& 1534~GeV\\
$\kappa$              		& -0.122			& $A_\lambda$	              & 1614~GeV		&$\kappa$              			&  -0.005		& $A_\lambda$	              & 1780~GeV\\
$\tan{\beta}$         		& 53.37			& $M_1$               		& -338.0~GeV	&$\tan{\beta}$         			& 58.54			& $M_1$               		& -706.3~GeV\\
$\mu_{\rm tot}$                	& 616.5~GeV	& $M_2$             		 	& 350.0~GeV	&$\mu_{\rm tot}$                		& 451.0~GeV	& $M_2$             		 	& 610.0~GeV\\
$m_B$					& 103.0~GeV	& $M_{\tilde{\mu}_L}$         & 506.3~GeV	&$m_B$						& 101.5~GeV	& $M_{\tilde{\mu}_L}$         & 528.8~GeV\\
$m_N$					& 604.3~GeV	& $M_{\tilde{\mu}_R}$		& 856.8~GeV	&$m_N$						& 437.1~GeV	& $M_{\tilde{\mu}_R}$		& 549.2~GeV\\
\hline
$m_{h_s}$                		& 95.92~GeV	&$m_{\tilde{\chi}_1^0}$   	& -337.2~GeV	&$m_{h_s}$                			& 96.1~GeV		&$m_{\tilde{\chi}_1^0}$   		& 436.8~GeV\\
$m_{h}$              			& 124.9~GeV	&$m_{\tilde{\chi}_2^0}$   	& -361.8~GeV	&$m_{h}$              			& 124.8~GeV	&$m_{\tilde{\chi}_2^0}$   		& 448.2~GeV\\
$m_{H}$                  		& 2412~GeV		&$m_{\tilde{\chi}_3^0}$   	& 603.8~GeV	&$m_{H}$                  			& 2377~GeV		&$m_{\tilde{\chi}_3^0}$   		& -462.3~GeV\\
$m_{A_s}$                 		& 797.7~GeV	&$m_{\tilde{\chi}_4^0}$   	& -635.9~GeV	&$m_{A_s}$                 		& 799.9~GeV	&$m_{\tilde{\chi}_4^0}$   		& 655.2~GeV\\
$m_{A_H}$                		& 2412~GeV		&$m_{\tilde{\chi}_5^0}$   	& 642.9~GeV	&$m_{A_H}$                		& 2377~GeV		&$m_{\tilde{\chi}_5^0}$   		& -712.0~GeV\\
$m_{H^\pm}$			& 2444~GeV		&$m_{\tilde{\mu}_L}$		& 512.5~GeV	&$m_{H^\pm}$				& 2404~GeV		&$m_{\tilde{\mu}_L}$			& 539.0~GeV\\
$m_{\tilde{\chi}_1^\pm}$	& 362.1~GeV	&$m_{\tilde{\mu}_R}$		& 857.9~GeV	&$m_{\tilde{\chi}_1^\pm}$	 	& 449.9~GeV	&$m_{\tilde{\mu}_R}$			& 554.1~GeV\\
$m_{\tilde{\chi}_2^\pm}$	& 645.1~GeV	&$m_{\tilde{\nu}_L}$		& 506.0~GeV	&$m_{\tilde{\chi}_2^\pm}$	 	& 655.5~GeV	&$m_{\tilde{\nu}_L}$			& 533.4~GeV\\
\hline
\multicolumn{2}{l}{$\Omega h^2 $}			&\multicolumn{2}{l|}{0.11}
&\multicolumn{2}{l}{$\Omega h^2 $}		&\multicolumn{2}{l}{0.10}\\
\multicolumn{2}{l}{$\sigma^{SI}_p$,  ~$\sigma^{SD}_n$}		&\multicolumn{2}{l|}{$8.6\times 10^{-47}{\rm ~cm^2}, ~1.3\times 10^{-42}{\rm ~~cm^2}$}
&\multicolumn{2}{l}{$\sigma^{SI}_p$, ~$\sigma^{SD}_n$}		&\multicolumn{2}{l}{$1.0\times 10^{-47}{\rm ~cm^2},  ~5.8\times 10^{-46}{\rm ~~cm^2}$}\\
\hline
\multicolumn{2}{l}{$a_{\mu}^{\rm SUSY} $,~$\chi^2_{\Delta_{a_\mu}}$}		&\multicolumn{2}{l|}{$2.59\times 10^{-9}$, ~0.04}
&\multicolumn{2}{l}{$a_{\mu}^{\rm SUSY} $,~$\chi^2_{\Delta_{a_\mu}}$}		&\multicolumn{2}{l}{$2.56\times 10^{-9}$, ~0.02} \\

\multicolumn{2}{l}{$a_{\mu}^{ \rm WHL}, ~a_{\mu}^{ \rm BHL}$}		
&\multicolumn{2}{l|}{$2.93\times 10^{-9}, ~~-1.68\times 10^{-10}$}
&\multicolumn{2}{l}{$a_{\mu}^{ \rm WHL} , ~a_{\mu}^{ \rm BHL} $}		
&\multicolumn{2}{l}{$2.71\times 10^{-9}, ~~-1.64\times 10^{-10}$}\\

\multicolumn{2}{l}{$ a_{\mu}^{\rm BHR} , ~a_{\mu}^{\rm BLR} $}		
&\multicolumn{2}{l|}{$1.20\times 10^{-10}, ~-2.24\times 10^{-10}$}
&\multicolumn{2}{l}{$a_{\mu}^{\rm BHR} , ~a_{\mu}^{ \rm BLR} $}		
&\multicolumn{2}{l}{$3.09\times 10^{-10}, ~-2.32\times 10^{-10}$}\\
\hline
\multicolumn{2}{l}{$\mu_{\gamma\gamma}, ~\mu_{b\bar{b}},~\Delta_{\gamma \gamma + b\bar{b}}$}		&\multicolumn{2}{l|}{0.22, ~0.13,~0.12}		
&\multicolumn{2}{l}{$\mu_{\gamma\gamma}, ~\mu_{b\bar{b}},~\Delta_{\gamma \gamma + b\bar{b}}$}	&\multicolumn{2}{l}{0.24, ~0.12,~0.0} \\
\multicolumn{2}{l}{$V_{h_s}^{S}, ~V_{h_s}^{SM}, ~V_{h_s}^{NSM}$}  & \multicolumn{2}{l|}{-0.930, ~0.367, ~$6.76\times 10^{-4}$}
& \multicolumn{2}{l}{$V_{h_s}^{S}, ~V_{h_s}^{SM}, ~V_{h_s}^{NSM}$}  & \multicolumn{2}{l}{-0.933, ~0.361, ~$9.66\times 10^{-4}$} \\
\multicolumn{2}{l}{$C_{h_s gg}, ~C_{h_s VV}, ~C_{h_s \gamma\gamma}, ~C_{h_s t \bar{t}}, ~C_{h_s b\bar{b}}$} 	& \multicolumn{2}{l|}{0.382, ~~0.367,~~~0.409,~~~0.367,~~0.331}
&\multicolumn{2}{l}{$C_{h_s gg}, ~C_{h_s VV}, ~C_{h_s \gamma\gamma}, ~C_{h_s t \bar{t}}, ~C_{h_s b\bar{b}}$} 	& \multicolumn{2}{l}{0.366,~~~0.351,~~~0.391,~~~0.351,~~0.291} \\
 \multicolumn{2}{l}{$C_{h gg}, ~~C_{h VV}, ~~C_{h \gamma\gamma}, ~~C_{h t \bar{t}}, ~~C_{h b\bar{b}}$} 			& \multicolumn{2}{l|}{0.376,~~~0.361, ~~0.402,~~~0.361,~~~0.304}
& \multicolumn{2}{l}{$C_{h gg}, ~~C_{h VV}, ~~C_{h \gamma\gamma}, ~~C_{h t \bar{t}}, ~~C_{h b\bar{b}}$} 		& \multicolumn{2}{l}{0.936,~~~0.933, ~~1.040,~~~0.933,~~~0.960} \\
\hline
\multicolumn{2}{l}{$N_{11}, ~N_{12}, ~N_{13}, ~N_{14}, ~N_{15}$}   &\multicolumn{2}{l|}{-0.994, ~-0.005,  ~-0.096, ~-0.050,  ~-0.000}
& \multicolumn{2}{l}{$N_{11}, ~N_{12}, ~N_{13}, ~N_{14}, ~N_{15}$}   &\multicolumn{2}{l}{0.004, ~~0.044, ~-0.111, ~~0.110, ~-0.987} \\
\multicolumn{2}{l}{$N_{21}, ~N_{22}, ~N_{23}, ~N_{24}, ~N_{25}$}   &\multicolumn{2}{l|}{0.007, ~~0.978, ~-0.182, ~~0.105, ~-0.001}
& \multicolumn{2}{l}{$N_{21}, ~N_{22}, ~N_{23}, ~N_{24}, ~N_{25}$}   &\multicolumn{2}{l}{-0.025,~~-0.280,~~~0.680,~~-0.657,~~-0.162} \\
\multicolumn{2}{l}{$N_{31}, ~N_{32}, ~N_{33}, ~N_{34}, ~N_{35}$}   &\multicolumn{2}{l|}{-0.001,~~-0.009,~~~0.024,~~-0.026,~~0.999}
& \multicolumn{2}{l}{$N_{31}, ~N_{32}, ~N_{33}, ~N_{34}, ~N_{35}$}   &\multicolumn{2}{l}{-0.122,~~-0.050,~~-0.701,~~-0.702, ~-0.002} \\
\multicolumn{2}{l}{$N_{41}, ~N_{42}, ~N_{43}, ~N_{44}, ~N_{45}$}   &\multicolumn{2}{l|}{-0.103,~~~0.055,~~~0.699,~~~0.706,~~~0.001}
& \multicolumn{2}{l}{$N_{41}, ~N_{42}, ~N_{43}, ~N_{44}, ~N_{45}$}   &\multicolumn{2}{l}{0.008,~~-0.958,~~-0.167,~~~0.234,~~~0.002} \\
\multicolumn{2}{l}{$N_{51}, ~N_{52}, ~N_{53}, ~N_{54}, ~N_{55}$}   &\multicolumn{2}{l|}{0.032,~~-0.203,~~-0.685,~~~0.698,~~~0.036}
& \multicolumn{2}{l}{$N_{51}, ~N_{52}, ~N_{53}, ~N_{54}, ~N_{55}$}   &\multicolumn{2}{l}{0.992,~~-0.006,~~-0.068,~~-0.106,~~-0.000} \\
\hline
\multicolumn{2}{l}{ Coannihilations }                         & \multicolumn{2}{l|}{Fractions [\%]} 			& \multicolumn{2}{l}{Coannihilations}                                       & \multicolumn{2}{l}{Fractions [\%]}                                      \\
\multicolumn{2}{l}{$\tilde{\chi}_2^0\tilde{\chi}_1^- \to d_i \bar{u}_i / \nu_{\ell_i} \ell_i^- / Z W^- / \gamma W^- $} 			& \multicolumn{2}{l|}{29.3/7.75/6.35/1.69}
& \multicolumn{2}{l}{$\tilde{\chi}_2^0\tilde{\chi}_1^- \to d_i \bar{u}_i / \nu_{\ell_i} \ell_i^- / h W^- / Z W^-  $} 				& \multicolumn{2}{l}{27.3/9.33/3.25/2.70}  \\

\multicolumn{2}{l}{$\tilde{\chi}_1^\pm \tilde{\chi}_1^\mp \to q_i \bar{q}_i / Z Z  /  W^- W^+ /\gamma Z $} 		& \multicolumn{2}{l|}{13.7/4.50/3.90/2.55}
& \multicolumn{2}{l}{$\tilde{\chi}_1^\pm \tilde{\chi}_1^\mp \to q_i \bar{q}_i / W^- W^+ / h Z / \gamma Z$} 		& \multicolumn{2}{l}{12.9/3.44/1.82/1.58}   \\

\multicolumn{2}{l}{$\tilde{\chi}_2^0 \tilde{\chi}_2^0 \to W^- W^+ $} 					& \multicolumn{2}{l|}{7.68}
& \multicolumn{2}{l}{$\tilde{\chi}_3^0\tilde{\chi}_1^- \to d_i \bar{u}_i / Z W^- $} 		& \multicolumn{2}{l}{6.63 / 1.10} \\  \hline

\multicolumn{2}{l}{ Decays }                         & \multicolumn{2}{l|}{Branching ratios [\%]} 		& \multicolumn{2}{l}{Decays}                                       & \multicolumn{2}{l}{Branching ratios [\%]}\\

\multicolumn{2}{l}{$\tilde{\chi}^0_2 \to \tilde{\chi}^0_1 Z^\ast$}      & \multicolumn{2}{l|}{100}
&\multicolumn{2}{l}{$\tilde{\chi}^0_2 \to \tilde{\chi}^0_1 Z^\ast$}      & \multicolumn{2}{l}{100} \\

\multicolumn{2}{l}{$\tilde{\chi}^0_3 \to \tilde{\chi}^\pm_1 W^\mp / \tilde{\chi}^0_2 h / \tilde{\chi}^0_1 Z / \tilde{\chi}^0_2 h_s$}      & \multicolumn{2}{l|}{64.7/24.4 /4.81 /3.64}
&\multicolumn{2}{l}{$\tilde{\chi}^0_3 \to \tilde{\chi}^\pm_1 W^{\mp\ast} /   \tilde{\chi}^0_{1,2} Z^\ast $}      & \multicolumn{2}{l}{48.5/47.1} \\

\multicolumn{2}{l}{$\tilde{\chi}^0_4 \to \tilde{\chi}^\pm_1 W^\mp / \tilde{\chi}^0_2 Z / \tilde{\chi}^0_1 h /  \tilde{\chi}^0_2 h$}      & \multicolumn{2}{l|}{61.5/27.0/7.29/1.41}
&\multicolumn{2}{l}{$\tilde{\chi}^0_4 \to \tilde{\chi}^\pm_1 W^\mp / \tilde{\chi}^0_3 Z / \tilde{\chi}^0_2 h/ \nu_{\mu} \tilde{\nu}_{\mu} /  \mu \tilde{\mu}_L $}      & \multicolumn{2}{l}{49.3/18.1/13.5/8.10/ 6.98 } \\

\multicolumn{2}{l}{$\tilde{\chi}^0_5 \to \tilde{\chi}^\pm_1 W^\mp / \tilde{\chi}^0_2 h / \tilde{\chi}^0_1 Z / \tilde{\chi}^0_2 h_s$}      & \multicolumn{2}{l|}{62.3/21.4/8.49/3.68}
&\multicolumn{2}{l}{$\tilde{\chi}^0_5 \to \tilde{\chi}^\pm_1 W^\mp / \mu \tilde{\mu}_R/ \tilde{\chi}^0_3 h / \mu \tilde{\mu}_L / \nu_{\mu} \tilde{\nu}_{\mu}$}      & \multicolumn{2}{l}{29.7/25.2/11.7/8.35/8.23} \\

\multicolumn{2}{l}{$\tilde{\chi}^+_1 \to \tilde{\chi}^0_1 W^{+\ast}$}      & \multicolumn{2}{l|}{100}
&\multicolumn{2}{l}{$\tilde{\chi}^+_1 \to \tilde{\chi}^0_1 W^{+\ast}$}      & \multicolumn{2}{l}{100}  \\

\multicolumn{2}{l}{$\tilde{\chi}^+_2 \to \tilde{\chi}^0_2 W^+ / \tilde{\chi}^+_1 Z / \tilde{\chi}^+_1 h / \tilde{\chi}^0_1 W^+$}      & \multicolumn{2}{l|}{31.6/30.1/23.0/10.4}
&\multicolumn{2}{l}{$\tilde{\chi}^+_2 \to \tilde{\chi}^0_{2,3} W^+ / \tilde{\chi}^+_1  Z / \tilde{\chi}^+_1 h/ \nu_{\mu} \tilde{\nu}_{\mu}/ \mu \tilde{\mu}_L$}      & \multicolumn{2}{l}{45.1/21.9/14.8/7.81/7.27}\\

\multicolumn{2}{l}{$\tilde{\mu}_L \to \nu_\mu \tilde{\chi}^\pm_1 / \mu \tilde{\chi}^0_2 / \mu \tilde{\chi}^0_1$}      & \multicolumn{2}{l|}{57.5/29.7/12.7}
&\multicolumn{2}{l}{$\tilde{\mu}_L \to \nu_\mu \tilde{\chi}^\pm_1 / \mu \tilde{\chi}^0_2 / \mu \tilde{\chi}^0_3 / \mu \tilde{\chi}^0_1$}      & \multicolumn{2}{l}{51.5/41.2/6.07/1.28}\\

\multicolumn{2}{l}{$\tilde{\mu}_R \to \mu \tilde{\chi}^0_1 / \mu \tilde{\chi}^0_4 /  \nu_{\mu} W^\pm$}      & \multicolumn{2}{l|}{99.0/0.37/0.22}
&\multicolumn{2}{l}{$\tilde{\mu}_R \to \mu \tilde{\chi}^0_3 / \mu \tilde{\chi}^0_2 / \nu_\mu \tilde{\chi}^\pm_1 $}      & \multicolumn{2}{l}{89.8/8.35/1.52}\\

\multicolumn{2}{l}{$\tilde{\nu}_{\mu} \to \mu \tilde{\chi}^\pm_1 / \nu_{\mu} \tilde{\chi}^0_2 / \nu_{\mu} \tilde{\chi}^0_1 $}      & \multicolumn{2}{l|}{59.1/28.7/12.1}
&\multicolumn{2}{l}{$\tilde{\nu}_{\mu} \to \mu \tilde{\chi}^\pm_1 / \nu_{\mu} \tilde{\chi}^0_2 / \nu_{\mu} \tilde{\chi}^0_1 $}      & \multicolumn{2}{l}{75.2/23.9/0.74}\\

\multicolumn{2}{l}{$h_s \to b \bar{b} / \tau^+ \tau^- / gg / c \bar{c} / W ^+W^{-\ast} / \gamma \gamma$}      & \multicolumn{2}{l|}{80.5/8.93/5.07/4.69/0.46/0.25}
&\multicolumn{2}{l}{$h_s \to b \bar{b} / \tau^+ \tau^- / gg / c \bar{c} / W ^+W^{-\ast} / \gamma \gamma$}      & \multicolumn{2}{l}{79.3/8.79/5.75/5.28/0.54/0.29}\\

\multicolumn{2}{l}{$h \to b \bar{b} / W ^+W^{-\ast} / \tau^+ \tau^- / gg / \gamma \gamma$}      & \multicolumn{2}{l|}{57.6/25.8/6.69/4.61/0.31}
&\multicolumn{2}{l}{$h \to b \bar{b} / W ^+W^{-\ast} / \tau^+ \tau^- / gg / \gamma \gamma$}      & \multicolumn{2}{l}{58.1/25.4/67.4/4.57/0.30}\\

\multicolumn{2}{l}{$H \to b \bar{b} /  \tilde{\chi}^\pm_1 \tilde{\chi}^\mp_2 /  \tau^+ \tau^-  /  \tilde{\chi}^0_2 \tilde{\chi}^0_{4,5}$}      & \multicolumn{2}{l|}{56.7/16.1/14.3/7.21}
&\multicolumn{2}{l}{$H \to b \bar{b} / \tau^+ \tau^- / \tilde{\chi}^\pm_1 \tilde{\chi}^\mp_2 / \tilde{\chi}^0_{2,3} \tilde{\chi}^0_4$}      & \multicolumn{2}{l}{61.9/14.7/12.2/5.89}\\

\multicolumn{2}{l}{$A_s \to b \bar{b} / \tilde{\chi}^\pm_1 \tilde{\chi}^\mp_1 / \tau^+ \tau^- / \tilde{\chi}^0_2 \tilde{\chi}^0_2 $}      & \multicolumn{2}{l|}{69.0/12.8/10.5/6.60}
&\multicolumn{2}{l}{$A_s \to b \bar{b} / \tau^+ \tau^- $}      & \multicolumn{2}{l}{86.6/13.2}\\

\multicolumn{2}{l}{$A_H \to b \bar{b} / \tilde{\chi}^\pm_1 \tilde{\chi}^\mp_2 / \tau^+ \tau^- / \tilde{\chi}^0_2 \tilde{\chi}^0_{4,5} $}      & \multicolumn{2}{l|}{56.7/15.8/14.3/7.21}
&\multicolumn{2}{l}{$A_H \to b \bar{b} / \tau^+ \tau^- / \tilde{\chi}^\pm_1 \tilde{\chi}^\mp_2 / \tilde{\chi}^0_{2,3} \tilde{\chi}^0_4$}      & \multicolumn{2}{l}{61.9/14.7/11.6/5.71}\\

\multicolumn{2}{l}{$H^+ \to t \bar{b} / \tau^+ \nu_{\tau} / \tilde{\chi}^0_2 \tilde{\chi}^+_2 / \tilde{\chi}^0_5 \tilde{\chi}^+_1 / \tilde{\chi}^0_4 \tilde{\chi}^+_1 $}      & \multicolumn{2}{l|}{55.7/15.6/8.84/8.32/7.2}
&\multicolumn{2}{l}{$H^+ \to t \bar{b} / \tau^+ \nu_{\tau} / \tilde{\chi}^0_2 \tilde{\chi}^+_2 / \tilde{\chi}^0_4 \tilde{\chi}^+_1 / \tilde{\chi}^0_3 \tilde{\chi}^+_2$}      & \multicolumn{2}{l}{60.9/16.1/7.00/6.69/6.04}\\
\hline

\multicolumn{2}{l}{$R$ value: 0.47}  & \multicolumn{2}{l|}{Signal Region: SR-A14 in Ref.~\cite{CMS:2017moi}}
& \multicolumn{2}{l}{$R$ value: 0.50}  & \multicolumn{2}{l}{Signal Region: SR-A44 in Ref.~\cite{CMS:2017moi}} \\

\multicolumn{2}{l}{$\chi_{tot}^2 \equiv \chi^2_{\Delta a_\mu} + \chi^2_{\gamma \gamma + b\bar{b}}$}  & \multicolumn{2}{l|}{0.16}
& \multicolumn{2}{l}{$\chi_{tot}^2 \equiv \chi^2_{\Delta a_\mu} + \chi^2_{\gamma \gamma + b\bar{b}}$}  & \multicolumn{2}{l}{0.02} \\
\hline \hline

\end{tabular}}
\caption{\label{Table6} Details of the best points for the Bino- and Singlino-dominated DM cases. Both of them can explain the $(g-2)_\mu$ anomaly and the diphoton and $b\bar{b}$ excesses at a $1\sigma$ level simultaneously, aligned with the LHC restrictions. They achieve the measured abundance primarily by co-annihilating with the Wino- and Higgsino-dominated electroweakinos, respectively. $d_i$, $u_i$, and $\ell_i$ appearing in the annihilation final state denote the $i$th generation of down-type quarks, up-type quarks, and leptons, respectively.}
\end{table}

We also examine the characteristics of the SM-like Higgs boson in a similar manner and present the results in Figs.~\ref{Fig6} and~\ref{Fig7}. These figures indicate that the couplings of $h$ are in agreement with those of the SM Higgs boson, with uncertainties of up to $10\%$, and that $h$ exhibits properties similar to those of $h_s$. However, there are two notable distinctions. Firstly, both $C_{h b \bar{b}}$ and $C_{h \gamma \gamma}$ take higher values compared to $C_{h t \bar{t}}$, with the latter surpassing $C_{h t \bar{t}}$ by approximately $10\%$. Secondly, $Br_{\rm SUSY} (h_s \to \gamma \gamma)$ and $Br_{\rm SUSY} (h_s \to b \bar{b})$ display a loose correlation and both vary within relatively broad ranges. Specifically,  ${\rm Br}(h \to b \bar{b})$ varies from $55.8 \%$ to $59.5 \%$, which is consistent with its SM prediction of $(57.7 \pm 1.8)\%$~\cite{LHCHiggsCrossSectionWorkingGroup:2013rie}, while  ${\rm Br} (h \to \gamma \gamma)$ ranges from $2.95 \times 10^{-3}$ to $3.13 \times 10^{-3}$, significantly exceeding  the SM prediction of $(2.28 \pm 0.11) \times 10^{-3}$~\cite{LHCHiggsCrossSectionWorkingGroup:2013rie}. We add that considering $C_{h g g} < 1$, the diphoton signal of $h$ remains consistent with its SM prediction.

Finally, we elucidate the physics underlying the observed excesses by providing detailed information on two benchmark points in Table~\ref{Table6}. These points represent the most favored solutions for the anomalies within the two types of DM scenarios, as indicated by stars in the earlier diagrams. Given that numerous characteristics of these points have been previously discussed, our focus now lies on highlighting the most significant signal at the LHC. Specifically, we concentrate on Higgsino-like and Wino-like electroweakino pair productions for points P1 and P2, respectively, leading to rich productions of $V_1 V_2 V_3^\ast V_4^\ast$ state from the sparticle decays, where $V_i$ with $i=1,2,3,4$ symbolizes either a W or Z boson. As suggested by the SRs A14 and A44 in Ref.~\cite{CMS:2017moi}, an optimal detection strategy for the production involves searching for events with three electrons or muons that include at least one opposite-sign same-flavor (OSSF) pair. The two SRs are distinct by the requirement for the invariant mass of the lepton pair to be less than $75~{\rm GeV}$ and greater than $105~{\rm GeV}$, respectively. Both points anticipate an R-value less than 1, implying their compatibility with restrictions from the LHC's search for sparticles~\cite{Cao:2021tuh}. Additionally, we make several observations regarding these two benchmark points:
 \begin{itemize}
 \item Considering that the analyses in Ref.~\cite{CMS:2017moi} utilized data from the initial year of LHC Run-2, accumulating $35.9 {~\rm fb^{-1}}$, re-analyzing the complete Run-2 dataset using the same strategy is anticipated to significantly enhance the LHC's sensitivity to these signals.
 \item For point P1's Wino-like and point P2's Higgsino-like electroweakino pair productions at the LHC, their ability to provide substantial evidence for supersymmetry is hindered by the  large DM mass. Specifically, these processes are effective in probing supersymmetry only when the DM is lighter than approximately $220~{\rm GeV}$ and $190~{\rm GeV}$, respectively, as demonstrated in Ref.~\cite{ATLAS:2021moa} through examinations of compressed sparticle mass spectrum cases.
 \item The slepton pair productions at both points prove ineffective for probing supersymmetry due to the complex decay chains involved with the sleptons, compounded by the substantial masses of the DM and the sleptons, as indicated in Refs.~\cite{ATLAS:2019lff, CMS:2020bfa}.
 \item As suggested in Ref.~\cite{Chakraborti:2021dli,Chakraborti:2021mbr,Chakraborti:2020vjp,Chakraborti:2021squ}, the electroweakino and slepton sectors of these points could be exhaustively explored at both high-luminosity LHC and a high-energy $e^+ e^-$ collider with $\sqrt{s} > 1~{\rm TeV}$.
 \item Researches expounded in Refs.~\cite{Drechsel:2018mgd,Biekotter:2019mib,Robens:2022zgk,Zarnecki:2023bod} underscore that future Higgs factories stand to meticulously scrutinize the properties of a Higgs boson with a mass of approximate 95 GeV.
 \end{itemize}
 
 \subsection{Impacts of the recent $125~{\rm GeV}$ Higgs data}
 
 \begin{figure}[t]
	\centering
	\includegraphics[width=0.6\linewidth]{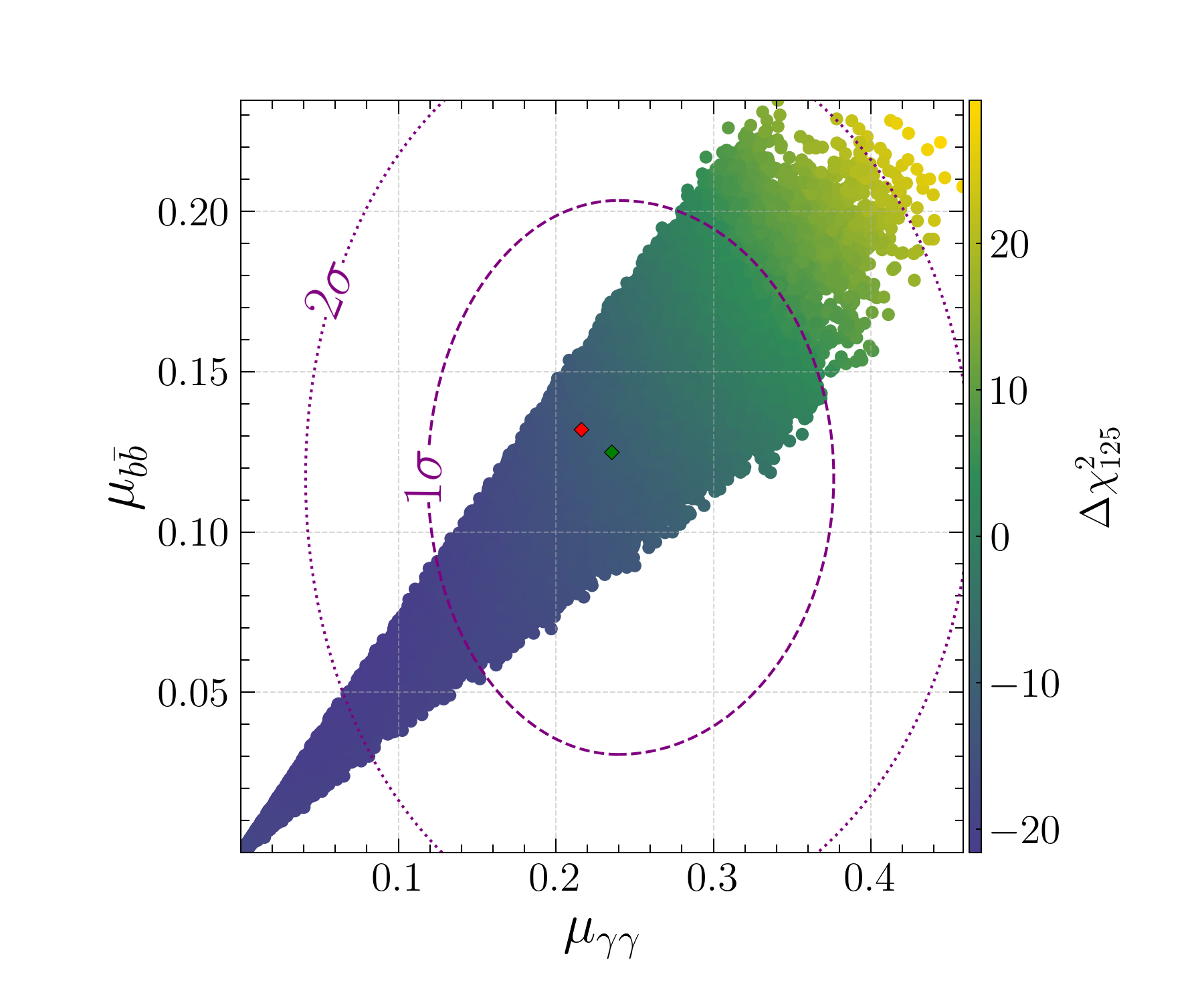}
\vspace{-0.3cm}
\caption{Projection of the parameter points from Fig.~\ref{Fig2} onto the $\mu_{\gamma \gamma}-\mu_{b\bar{b}}$ plane. The color bar represents $\Delta \chi^2_{125}$, defined as the difference between $\chi^2_{125}$ and $\chi^2_{\rm SM, 125}$: $\Delta \chi^2_{125} \equiv \chi^2_{125} - \chi^2_{\rm SM, 125}$. All fits for the 125~{\rm GeV} Higgs data were performed using \texttt{HiggsTools-1.2}, yielding  $\chi^2_{\rm SM, 125}=176.3$ for a fully SM-like Higgs boson with an assumed mass uncertainty of 3~{\rm GeV} (see text for details). The benchmark points P1 and P2 from Table~\ref{Table6} are indicated by red and green dots, respectively. \label{Fig8}}
\end{figure}

The ongoing operation of the LHC has led to a substantial accumulation of 125 GeV Higgs boson data, enabling increasingly precise measurements of its properties~\cite{ATLAS:2022vkf,CMS:2022dwd,ATLAS:2024fkg}. These enhanced precision measurements provide an opportunity to establish more stringent constraints on theories beyond the SM. To leverage these new measurements, we refined the Higgs data fitting in Sec.~\ref{Section-analysis} using \texttt{HiggsTools-1.2}, the latest version of \texttt{HiggsTools}~\cite{Bahl:2022igd}, to reassess previously scanned parameter points. This updated version marks a significant advancement over the \texttt{HiggsSignals-2.6.2} by incorporating 22 additional physical observables, thereby increasing the total number of observables available for fitting Higgs properties to 129~\cite{Bahl:2022igd}.

In our study, we initially specified particular parameter values: $\tan \beta =20$ and $\lambda = \kappa = 10^{-4}$. Additionally, we set all mass-dimensional parameters in the Higgs, sfermion, and gaugino sectors to 2 TeV, with $A_t$ being the sole exception. Through meticulous adjustment of $A_t$, we successfully reproduced a Higgs particle with a mass of approximately $125.1~{\rm GeV}$. This particle shares identical features with the SM Higgs boson, with the crucial difference that its mass is derived from sophisticate theoretical calculations rather than being an input parameter. Upon incorporating a mass uncertainty of 3 GeV into our analysis, the Higgs data fitting process  yielded a $\chi^2$ value of approximately 176.3, denoted as $\chi^2_{\rm SM, 125} \simeq 176.3$.

To evaluate how recent Higgs data influence the theoretical interpretation of the observed excesses, we mapped the parameter points from Fig.~\ref{Fig2} onto the $\mu_{\gamma \gamma}-\mu_{b \bar{b}}$ plane in Fig.~\ref{Fig8}, where the color bar indicates the difference between their $\chi^2$  values (denoted as $\chi^2_{125}$) in the $125~{\rm GeV}$ Higgs data fit using \texttt{HiggsTools-1.2} and $\chi^2_{\rm SM, 125}$. Our analysis demonstrates strong compatibility between these parameter points and the current Higgs data, as evidenced by the following observations:
\begin{itemize}
\item The benchmark points P1 and P2 presented in Table~\ref{Table6} exhibit $\chi^2_{125}$ values of 164.7 and 165.9 respectively, both falling below $\chi^2_{\rm SM, 125}$.
\item Within the experimentally acceptable range defined by $\chi^2_{125} \lesssim \chi^2_{\rm SM, 125} + 6.18$~\cite{Muhlleitner:2020wwk}, the signal strengths $\mu_{\gamma \gamma}$ and $\mu_{b \bar{b}}$ can achieve maximum values of 0.37 and 0.21, respectively.
\end{itemize}
It is noteworthy that $\chi^2_{\rm SM, 125}/\nu \simeq 1.38$ significantly exceeds $1 + \sqrt{2/\nu} = 1.125$, where $\nu \equiv  n_{\rm obs} - n_{\rm param}$ denotes the degrees of freedom with $n_{\rm obs} = 129$ representing the number of physical observables in  \texttt{HiggsTools} and $n_{\rm param} = 1$ corresponding to the Higgs boson mass in the SM~\cite{Bechtle:2012jw}. As emphasized in Ref.~\cite{Bahl:2022igd}, this discrepancy indicates substantial systematic uncertainties in the data incorporated within the \texttt{HiggsTools}.

\section{Conclusion} \label{conclusion}
Recently, FermiLab has updated its measurement of the muon anomalous magnetic moment based on Run-2 and Run-3 data, revealing a significant deviation from the current SM prediction at a confidence level of $5.1\sigma$. Despite uncertainties in calculating leading-order hadronic contributions, this observation likely indicates the presence of new physics. Moreover, both the CMS and ATLAS collaborations have recently reported an excess in the diphoton invariant mass distribution at $m_{\gamma \gamma} \simeq 95.4~{\rm GeV}$ with compatible signal strengths. By combining these novel findings with a similar excess previously reported by CMS, the local significance has increased to $3.1\sigma$. Additionally, it is noteworthy that the invariant mass of the diphoton signal coincides with that of the observed $b\bar{b}$ excess at the LEP, suggesting they may originate from a same CP-even Higgs boson with its mass around $95.4~{\rm GeV}$.

The GNMSSM stands out for its theoretical advantages and a versatile parameter space, which enables a comprehensive exploration of Higgs and DM phenomenology. In this study, we apply the GNMSSM to simultaneously explain the muon $g-2$ anomaly, as well as the diphoton and $b\bar{b}$ excesses. We demonstrate that the anomaly can be entirely attributed to the loops involving muon-smuon-neutralino and muon-sneutrino-chargino interactions. By employing the mass insertion approximation of $a_\mu^{\rm SUSY}$, we find a preference for a large $\tan \beta$ value and a sub-TeV SUSY scale. Additionally, we address the excesses by considering resonant productions of the singlet-dominated scalar $h_s$. Our derivation of approximate formulae for signal strengths highlights how crucial input parameters such as $\lambda$, $\tan \beta$, $\mu_{\text{tot}}$, $A_\lambda$, and $m_N$ interact with each other. Our conclusion reveals that the unified explanations favor a moderately large SM Higgs field component in $h_s$, with a suppressed $h_s b \bar{b}$ coupling relative to $h_s t \bar{t}$ coupling. Small deviations in $C_{h_s g g}$ and $C_{h_s \gamma \gamma}$ from $C_{h_s t \bar{t}}$ enhance the theory's ability to explain the excesses.

To comprehensively investigate the impact of experimental restrictions on the unified explanation, encompassing those from Higgs physics, B-physics, DM physics, and collider searches for sparticles, we undertake an exploration across the GNMSSM parameter space. Our analysis yields the following key findings:
\begin{itemize}
\item
The GNMSSM can simultaneously account for the muon $(g-2)$ anomaly and the diphoton and $b\bar{b}$ excesses at a $2\sigma$ level within a wide range of parameters without conflicting with constraints from the SM-like Higgs data fit, the B-physics measurements, the Planck and LZ experiments, the vacuum stability considerations, as well as the LHC's searches for sparticles and extra Higgs bosons.

\item
DM physics significantly influences the unified explanation through the crucial parameter $\mu_{\rm tot}$. Specifically, the preferred samples are classified into either the Bino-DM scenario or Singlino-DM scenario based on their dominant DM component. They achieve the observed relic abundance primarily by co-annihilating with Wino-like and Higgsino-like electroweakinos respectively, contributing to $26\%$ and $74\%$ of the total Bayesian evidence. The Bino-DM scenario favors $\mu_{\rm tot} \gtrsim 600~{\rm GeV}$ mainly due to restrictions imposed by the LZ experiment, whereas in Singlino-DM scenario $\mu_{tot} \simeq |m_{\tilde{\chi}_1^0}|$ can be as low as $400~{\rm GeV}$. As indicated in this study, the difference has profound implications in $\delta a_\mu^{\rm SUSY}$, $\mu_{\gamma \gamma}$, and $\mu_{b \bar{b}}$.

\item
The ``WHL" contribution dominates $a_\mu^{\rm SUSY}$ calculation. It sets upper bounds on $\mu_{\rm tot}$, $M_2$ and $M_{\tilde{\mu}_L}$ in explaining the muon g-2 anomaly.
Additionally, the ``BHL" and ``BLR" contributions consistently counteract other terms when considering negative $M_1$ values, which are preferred in Bino-DM case to suppress DM-nucleon scattering.

\item
Constraints arising from the LHC's searches for sparticles slightly narrow down the parameter regions in the Bino-DM case but have minimal impacts on the unified explanation in the Singlino-DM case due to our ad hoc assumption regarding the range of $\mu_{tot}$, i.e., $\mu_{tot} \geq 400~{\rm GeV}$. Incorporating the LHC restrictions sets lower bounds on masses of the sparticles involved in the explanation, such as neutralinos, charginos, and muon-flavored sleptons.

\end{itemize}

Before concluding our study, we believe it is important to share the following remarks about our findings:
\begin{itemize}
\item In formulating the likelihood function, we have adopted overly idealistic assumptions regarding the prolific generation of diphoton and $b\bar{b}$ events at colliders. Furthermore, considering that increasingly comprehensive analyses of the experimental data are being carried out, it is likely that we might have underestimated the impact of the LHC's search for SUSY in excluding the parameter space of the GNMSSM. Consequently, although the results presented here provide valuable insights into the fundamental physics of the observed excesses, they do not possess strict statistical significance. For instance,  $\chi^2_{tot}= 0.02$ for the optimal point P2 in Table~\ref{Table6} merely indicates that the GNMSSM can predict the central values of the excesses, and it outperforms the SM, which yields $\chi^2_{tot}= 40.1$ in accounting for the observed excesses. It should not be overinterpreted.

   We remind that a more realistic methodology for constructing the likelihood function can be found in Ref.~\cite{GAMBIT:2017snp}, which addresses the global fitting of GUT-scale SUSY models to experimental data.

\item As demonstrated by this study, addressing the diphoton excess, the $b\bar{b}$ excess, and the muon g-2 anomaly demands a considerable $V_{h_s}^{\rm SM}$, an appropriate cancellation between $V_{h_s}^{\rm SM}$ and $V_{h_s}^{\rm NSM} \tan \beta$, and, depending on the value of $\tan \beta$, the presence of moderately light electroweakinos and muon-type sleptons, respectively.   This unified interpretation integrates these distinct requirements to delineate specific GNMSSM parameter configurations.  Some of these constraints would be alleviated if any excess diminishes, thereby expanding the permissable parameter space.   For example, if future theoretical and experimental advancements reveal that $\Delta a_\mu$ is significantly smaller than its current estimation, the results could be accommodated simply by increasing the sparticle mass values and in this case, moderately light sparticles are no longer necessary. As shown in this study, such adjustments in the GNMSSM's parameter space would not impact the explanations for the other anomalies and the constraints from the LHC's search for SUSY.

\item The muon g-2 anomaly has been comprehensively investigated within the $h_1$ and $h_2$ scenarios of the GNMSSM~\cite{Cao:2022ovk,Cao:2022chy}, where the SM-like Higgs boson $h$ corresponds to the lightest and next-to-lightest CP-even Higgs bosons, respectively.  It has been observed that the $h_1$ scenario provides a much broader parameter space compared to those of the MSSM and the $Z_3$-NMSSM for explaining the anomaly~\cite{Cao:2022ovk}.  This is attributed to more relaxed constraints from DM experiments and the LHC's search for SUSY. In contrast, the $h_2$ scenario has difficulty accounting for the anomaly as its parameter space is characterized by $\mu_{tot} \lesssim 500~{\rm GeV}$ and $\tan \beta \lesssim 30$, with fixed $A_\lambda = 2~{\rm TeV}$ and $\mu^\prime = 0$~\cite{Cao:2022chy}. This parameter configuration tends to result in a relatively low $ {\cal M}^2_{S, 23} $ in Eq.~\ref{Mass-CP-even-Higgs}, thereby avoiding overly large singlet-doublet Higgs mixing, which is more in line with the observed Higgs data.  However, as indicated by the study in Ref.~\cite{Cao:2022chy}, it faces strict limitations from the LHC's searches for SUSY. Our study improves upon Ref.~\cite{Cao:2022chy} by choosing the field masses $m_A$, $m_B$, and $m_C$ (instead of the soft-breaking masses in the original Lagrangian adopted in Ref.~\cite{Cao:2022chy}) as theoretical inputs and simultaneously allowing $A_\lambda$ to vary within reasonable ranges during the parameter scan of the GNMSSM, as detailed in Table~\ref{tab:scan}.  We draw conclusions that significantly differ from those of Ref.~\cite{Cao:2022chy}: the GNMSSM can still explain the muon g-2 anomaly across a vast parameter space that complies with experimental constraints. The key difference stems from the fact that our refined scanning strategy affects the posterior PDF of the samples, particularly showing that a considerable proportion of samples suggest $\tan \beta > 30$ and $\mu_{tot} > 300~{\rm GeV}$. This case demands a relatively small $\lambda$ and a moderate cancellation between different contributions to $ {\cal M}^2_{S, 23} $, effectively reducing its size by typically $10\%$. Such a case was neglected in the study of Ref.~\cite{Cao:2022chy}. Indeed, it is the limitations in the analysis of Ref.~\cite{Cao:2022chy} that motivated us to extend our previous study of the diphoton and $b\bar{b}$ excesses in the GNMSSM, as elaborated in Ref.~\cite{Cao:2023gkc}, to also incorporate the muon g-2 anomaly. 
 
\item  We note that the three anomalies were also investigated in Ref.~\cite{Ellwanger:2024txc} within the framework of the $Z_3$-NMSSM, a special implementation of the GNMSSM~\cite{Ellwanger:2009dp}.  That study was conducted later than ours and it was found that these anomalies could be accounted for in a parameter space characterized by $\lambda \gtrsim 0.2$, $\tan \beta \lesssim 20$, and the presence of at least one electroweakino or slepton lighter than approximately $200~{\rm GeV}$. These findings are significantly different from our current results.

The disparity in the two findings can be mainly attributed to our distinct research objectives. As discussed in Sec.~\ref{Section-analysis}, this study specifically focused on examining whether the GNMSSM could explain these three anomalies under various experimental constraints, rather than carrying out a comprehensive global fitting analysis.        We deliberately excluded the parameter regions discussed in Ref.~\cite{Ellwanger:2024txc} based on the following two crucial considerations:
\begin{itemize}
\item As elaborated below Eq.~(\ref{Approximation-relations-2}), the condition  $\mu_{tot} - \bar{A}_\lambda \sin \beta \cos \beta \simeq 10~{\rm GeV}$ is necessary for the substantial $\lambda$ case to account for the $\gamma \gamma$ and $b \bar{b}$ excesses.  This condition demands a strong cancellation between two independent terms  $\mu_{tot}$ and $\bar{A}_\lambda \sin \beta \cos \beta$, typically resulting in a net magnitude reduction by one order. Such fine-tuning can significantly suppress the posterior PDF considered in this study, rendering this explanation statistically less probable.

\item In scenarios where $\tan \beta \lesssim 20$, moderately light sparticles are needed to explain the muon $g-2$ anomaly. However, as demonstrated in Ref.~\cite{Cao:2022chy},  this scenario is subject to strict limitations from LHC supersymmetry searches, making viable solutions extremely rare. To appropriately account for LHC constraints in these cases, dedicated Monte Carlo simulations of the electroweakino and slepton production processes using CheckMATE are essential~\cite{Cao:2022ovk}.  This requirement stems from the fact that the capabilities of the code SModelS in this domain are significantly restricted when analyzing numerous realistic SUSY scenarios, especially in cases involving long decay chains of heavy sparticles, due to both its database limitations and strict usage prerequisites. While many previous studies, including that in Ref.~\cite{Ellwanger:2024txc}, overlooked this conclusion, our earlier research demonstrated that the SModelS could effectively exclude parameter points in only a small fraction of cases for both the  $Z_3$-NMSSM~\cite{Cao:2022htd} and the GNMSSM~\cite{Cao:2022ovk,Cao:2022chy}.  
\end{itemize}
    
\end{itemize}

\section*{Acknowledgements}

We thank Zehan Li  for helpful discussions about calculating $\chi^2$ in Higgs data fit.This work is supported by the National Natural Science Foundation of China (NNSFC) under grant No. 12075076.


\bibliographystyle{CitationStyle}
\bibliography{LianRef}

\end{document}